\pdfoutput=1
\documentclass[12pt,preprint]{aastex}
\usepackage{natbib}
\newcommand{\Msun}{M{$_{\odot}$}}

\newcommand{\Ha}{\mbox{H$\alpha$}}
\newcommand{\ha}{\mbox{H$\alpha$}}
\newcommand{\sii}{[S{\sc ii}]}
\newcommand{\mh}{H$_2$}
\newcommand{\feii}{[Fe{\sc ii}]}
\newcommand{\Ks}{K$_S$}
\newcommand{\um}{$\mu$m}

\slugcomment{Accepted for publication in the The Astrophysical Journal}

\shorttitle{L1340}
\shortauthors{Walawender et al.}

\begin{document}

\title{Protostellar Outflows in L1340}
\author{Josh Walawender}
\affil{W. M. Keck Observatory, 65-1120 Mamalahoa Hwy, Kamuela, HI 96743}
\email{jmwalawender@keck.hawaii.edu}
\author{Grace Wolf-Chase}
\affil{Astronomy Department, Adler Planetarium, 1300 South Lake Shore Drive, Chicago, IL 60605, USA}
\author{Michael Smutko}
\affil{Center for Interdisciplinary Exploration and Research in Astrophysics (CIERA) \& Department of Physics \& Astronomy, Northwestern University, 2145 Sheridan Road, Evanston, IL 60208, USA}
\author{JoAnn O’Linger-Luscusk.}
\affil{California Institute of Technology, 1200 E California Blvd, Pasadena, CA 91125}
\and
\author{Gerald Moriarty-Schieven}
\affil{National Research Council - Herzberg Astronomy \& Astrophysics, 5017 West Saanich Road, Victoria, BC, V9E 2E7, Canada}

\begin{abstract}
We have searched the L1340 A, B, and C clouds for shocks from protostellar outflows using the \mh{} 2.122\,\um{} near-IR line as a shock tracer.  Substantial outflow activity has been found in each of the three regions of the cloud (L1340\,A, L1340\,B, \& L1340\,C).  We find 42 distinct shock complexes (16 in L1340\,A, 11 in L1340\,B, and 15 in L1340\,C).  We were able to link 17 of those shock complexes in to 12 distinct outflows and identify candidate source stars for each.  We examine the properties (A$_{V}$, T$_{bol}$, and L$_{bol}$) of the source protostars and compare that to the properties of the general population of Class\,0/I and flat SED protostars and find that there is an indication, albeit at low statistical significance, that the outflow driving protostars are drawn from a population with lower A$_{V}$, higher L$_{bol}$, and lower T$_{bol}$ than the general population of protostars.
\end{abstract}

\section{Introduction}

Young stars interact with their parent molecular cloud through the action of protostellar outflows.  These magnetohydrodynamically driven outflows are launched at speeds of order 100\,$km/s$ as a byproduct of the accretion process.  This high velocity gas shocks against the surrounding material and is detected via shock tracers in the optical (such as \ha{} and \sii{} in which case the shocks are called Herbig-Haro or HH objects) or in the infrared (typically via \mh{} or \feii{} emission).  These outflows affect the parent cloud's dynamics by opening cavities (e.g. \citealt{Quillen05}) or by driving turbulence (e.g. \citealt{MieschBally94}, \citealt{Bally99}, \citealt{ArceGoodman2002}, \citealt{Walawender2005}).

We have surveyed the L1340\,A, B, and C clouds in the near-IR to look for shocks from protostellar outflows as revealed by their \mh{} 2.12\micron{} S(1$-$0) emission.  We found 42 shock complexes (MHO\,2925 to 2966 in the MHO catalog hosted by the University of Kent\footnote{\url{http://astro.kent.ac.uk/~df/MHCat/}} and initially published in \citealt{Davis2010}).

LDN\,1340 (L1340 hereafter) is an intermediate mass (mid-B and lower luminosity) star forming region in Cassiopeia.  It was number 1340 of 1802 in \cite{Lynds1962} catalog of dark nebulae.  \cite{DorschnerGurtler1963} catalogued a reflection nebula (DG\,9) in the dark cloud.  Later, \cite{Cohen1980} examined Palomar Sky Survey images and found three "Red Nebulous Objects" (RNO 7, 8, and 9) in the area.

The first detailed study of the area was carried out by \cite{Kun1994} who mapped the region in $^{13}$CO and C$^{18}$O and found a molecular mass of about 1300\,\Msun{}.  Based on their molecular maps, \cite{Kun1994} divided the L1340 cloud in to three sub-regions:  L1340\,A, B, and C corresponding roughly to the areas surrounding RNO\,7, 8, and 9 respectively.  

\cite{Yonekura1997} included L1340 in their $^{13}$CO survey of molecular clouds toward Cepheus and Cassiopeia and found a mass of 1200\,\Msun{}.  \cite{Kun2003} observed L1340 in NH$_3$ and found 10 dense cores with a total mass of about 80\,\Msun{}.  \cite{Juvela2012} included L1340 in their Herschel SPIRE observations and described the morphology of the cloud as filamentary rather than cometary.

\cite{Kun2011} performed photometric and spectroscopic monitoring of the HA11 star (using the nomenclature of \citealt{Kun1994}) and found it to be an eruptive young star, though not easily falling in to classification as either a FUor or EXor type.  \cite{Kun2014} discovered three additional candidate eruptive young stars in L1340: IRAS\,02224+7227 (discussed in \S\ref{MHO2925}), 2MASS\,02263797+7304575, and\,2MASS 02325605+7246055 (discussed in \S\ref{MHO2962}).

Two previous optical/near-IR surveys have been done in this region:  \cite{Kumar2002} and \cite{Magakian2003}.  \cite{Kumar2002} found three Herbig-Haro (HH) objects in L1340\,A.  \cite{Magakian2003} found two additional HH objects in L1340\,A and also catalogued 14 \ha{} emission line stars (many overlapping with the list compiled by \citealt{Kun1994}).

\begin{figure}[!htb]
\includegraphics[width=1.0\textwidth]{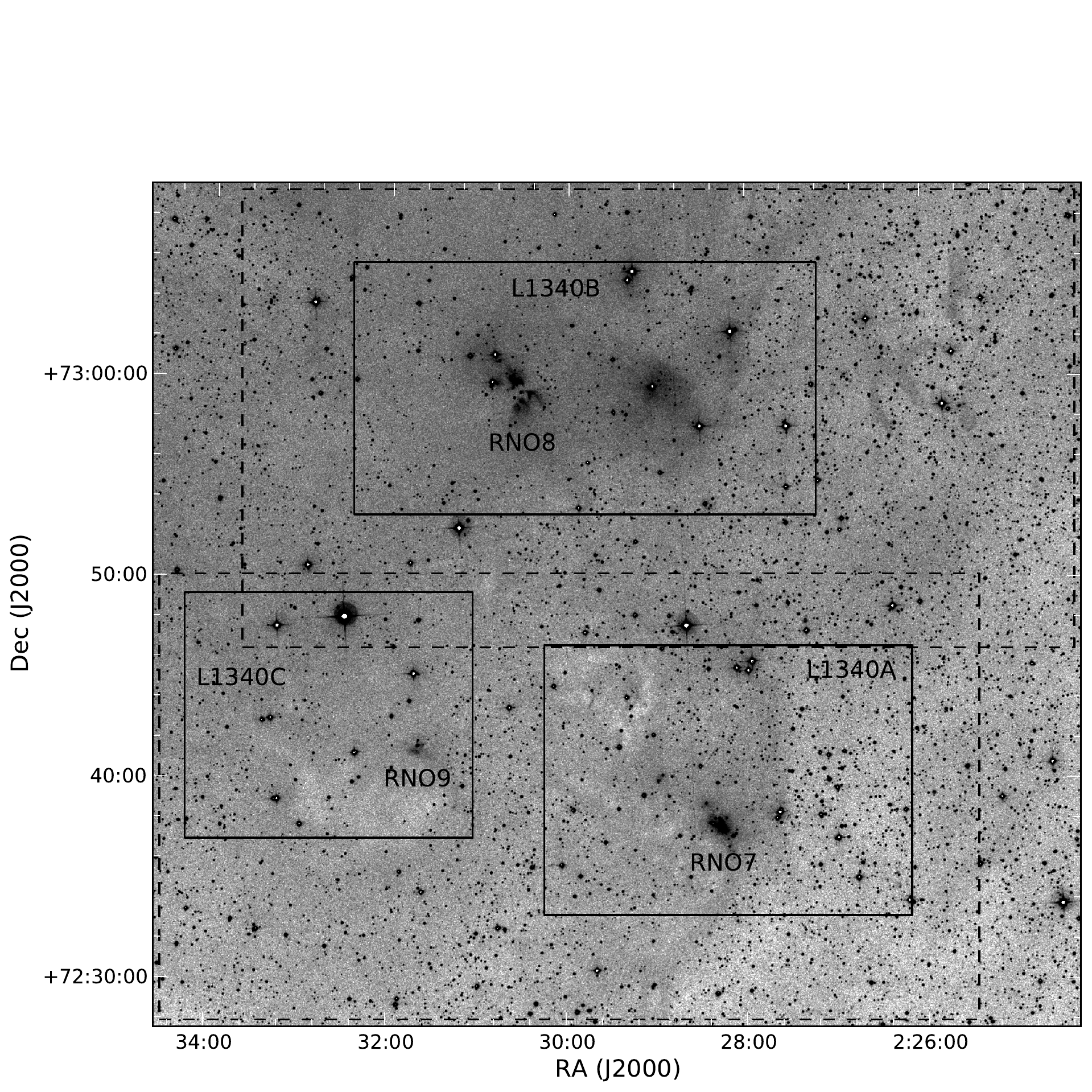}
\caption{The POSS R image of the L1340 cloud region.  The three RNO objects of \cite{Cohen1980} are labeled and the three regions of the cloud (L1340\,A, L1340\,B, and L1340\,C) are also marked.  Boxes with dashed outlines indicate the rough coverage of our WIRCam \mh{} images while boxes with solid outlines indicate the positions of Figs.\ \ref{FigL1340AOverview}, \ref{FigL1340BOverview}, \& \ref{FigL1340COverview} discussed later in the text.}
\label{FigL1340Overview}
\end{figure}

\cite{Kun2016Opt} used optical spectroscopy and infrared and optical photometry to examine the protostellar population in L1340 and mapped the extinction in the region based on SDSS photometry.  In addition, they use spectral classification of two ZAMS stars associated with the cloud to estimate a distance of $825^{+110}_{-80}$\,pc.

Using infrared photometry, \cite{Kun2016IR} identified and classified 170 Class\,II, 27 flat SED, and 45 Class\,0/I protostars in L1340.  They estimated the extinction to each using the results of \cite{Kun2016Opt} and derived bolometric temperatures and luminosities based on the spectral energy distribution (SED) of each protostar.

In the following sections, we describe our observations (\S\ref{observations}) and give detailed descriptions of the outflows in each of the A (\S\ref{L1340A}), B (\S\ref{L1340B}, and C (\S\ref{L1340C}) clouds.  In \S\ref{discussion} we compare L1340 with our prior observations of Barnard\,1 (\S\ref{B1}) and compare the properties of the outflows we have identified in L1340 with the properties of the protostars which drive them (\S\ref{properties}).  Our results are summarized in section \ref{summary}.

\section{Observations}\label{observations}

Data for this paper were taken during the 2006B and 2007B semesters with the Canada France Hawaii Telescope (CFHT) on Mauna Kea on the island of Hawai'i.  The instrument used was the Wide-field InfraRed Camera (WIRCam; \citealt{Pudget2004}) which consists of a mosaic of four HAWAII2-RG detectors, each containing 2048 $\times$ 2048 pixels, with a sampling of 0.3 arc second per pixel on the sky.  The data were pipeline processed by 'I'iwi, the standard WIRCam pipeline at Traitement \'{E}l\'{e}mentaire, R\'{e}duction et Analyse des PIXels (TERAPIX) which is located at Institut d'Astrophysique de Paris (IAP).


Our 2006B data cover the northern half of the cloud (L1340\,B) in \mh{}, H, and \Ks{}, while our 2007B data cover the southern half of the cloud (L1340\,A and L1340\,C) in \mh{} and all of the cloud in J, H, \& \Ks{} (see Fig.\ \ref{FigL1340Overview}).  The total integration time in \mh{} is 54 minutes at each position.  The integration times for the broadband filters varies depending on the position as some frames were rejected from the stacks due to poor image quality, but the minimum integration times were 3.5 minutes in J, 3.3 minutes in H, and 3.67 minutes in \Ks{}.  The seeing values in the final stacked images (as reported by the pipeline software) range from 0.75\arcsec{} to 1.55\arcsec{} FWHM.  The total area covered is about 1700 square arcminutes (0.47 square degrees) in each filter.

WIRCam images in all filters were registered to one another using their world coordinate system and the \mh{} images of the two halves of the cloud were stitched together using the MSCRED package in the Image Reduction and Analysis Facility\footnotemark (IRAF).  In addition, IRAF was used to generate difference images (\mh{} $-$ \Ks{}) after application of a gaussian blur (to more closely match PSF sizes) and a scaling factor (which was estimated from the flux of stars and reflection nebulae common to both images).

\footnotetext{IRAF is distributed by the National Optical Astronomy Observatories, which are operated by the Association of Universities for Research in Astronomy, Inc., under cooperative agreement with the National Science Foundation.}

In addition, \mh{} and \Ks{} images of a sub-region of the L1340\,B cloud were obtained at the ARC 3.5 meter telescope at Apache Point Observatory in New Mexico using the Near-Infrared Camera and Fabry-Perot Spectrometer (NICFPS; \citealt{Hearty2004}) instrument on the nights of 19 September 2005 and 9-13 January 2006.  NICFPS uses a 1024 $\times$ 1024 pixel Hawaii-1RG detector which has a pixel scale of 0.273 arcseconds per pixel on the sky.  The NICFPS images were processed and then registered with the WIRCam images using IRAF.  The NICFPS data cover about 221 square arcminutes (0.06 square degrees) in both \mh{} and \Ks{}.

\section{Results}\label{Results}

We find 42 distinct shock complexes (16 in L1340\,A, 11 in L1340\,B, and 15 in L1340\,C).  Each shock complex is given an MHO designation \citep{Davis2010} which is listed in Table \ref{TableShocks} and each is discussed below.  We group the discussions by region:  L1340\,A is covered in \S\ref{L1340A}, L1340\,B in \S\ref{L1340B}, and L1340\,C in \S\ref{L1340C}.

\subsection{L1340A}\label{L1340A}


\subsubsection{A 3.7\,pc Long Outflow Lobe:  MHO\,2925, MHO\,2937, \& HH 487}
\label{MHO2925}

\begin{figure}[!htb]
\includegraphics[width=1.0\textwidth]{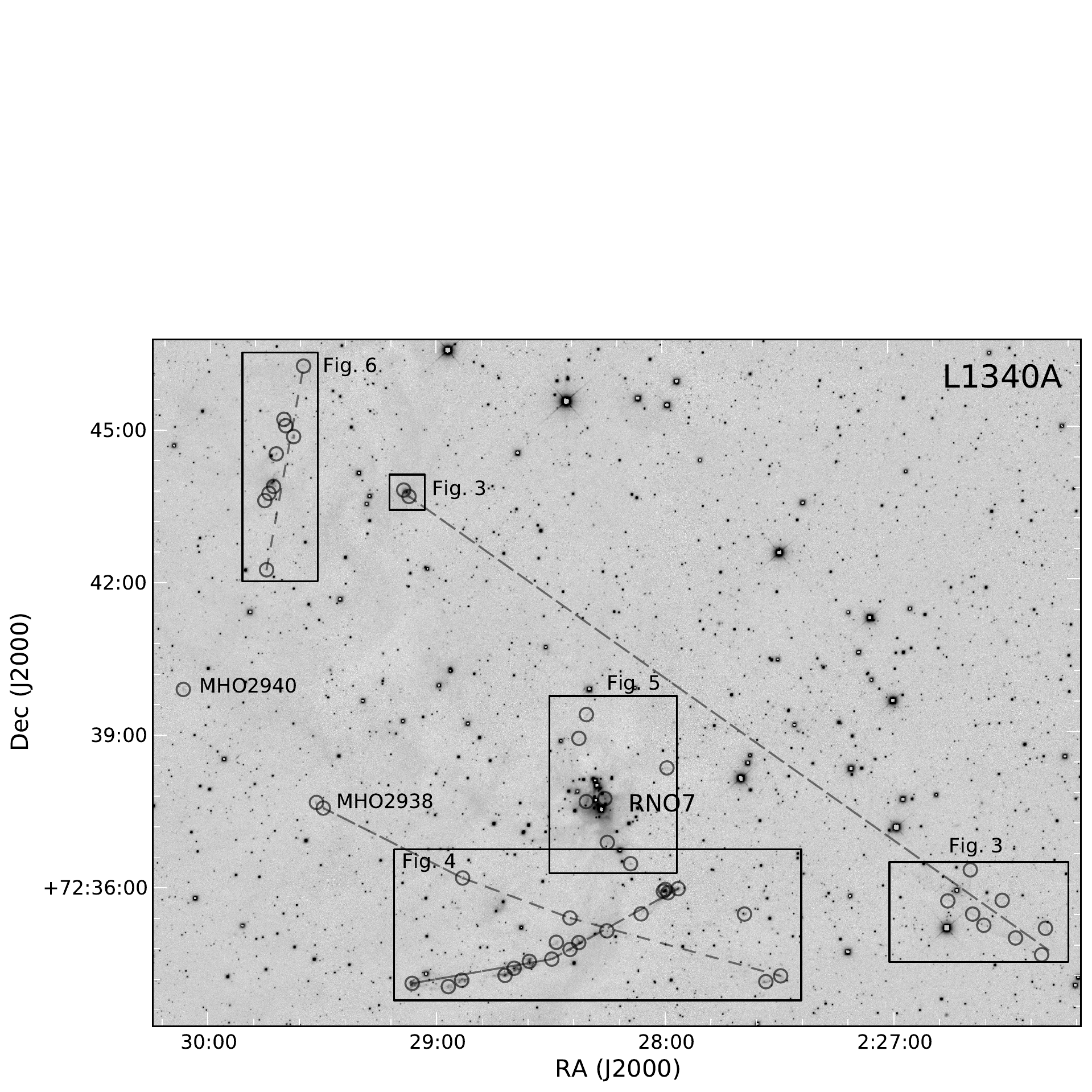}
\caption{An \mh{} image of the L1340\,A region.  Circles mark the positions of \mh{} shocks and boxes indicate the locations of subsequent figures in the text.}
\label{FigL1340AOverview}
\end{figure}

\begin{figure}[!htb]
\includegraphics[width=1.0\textwidth]{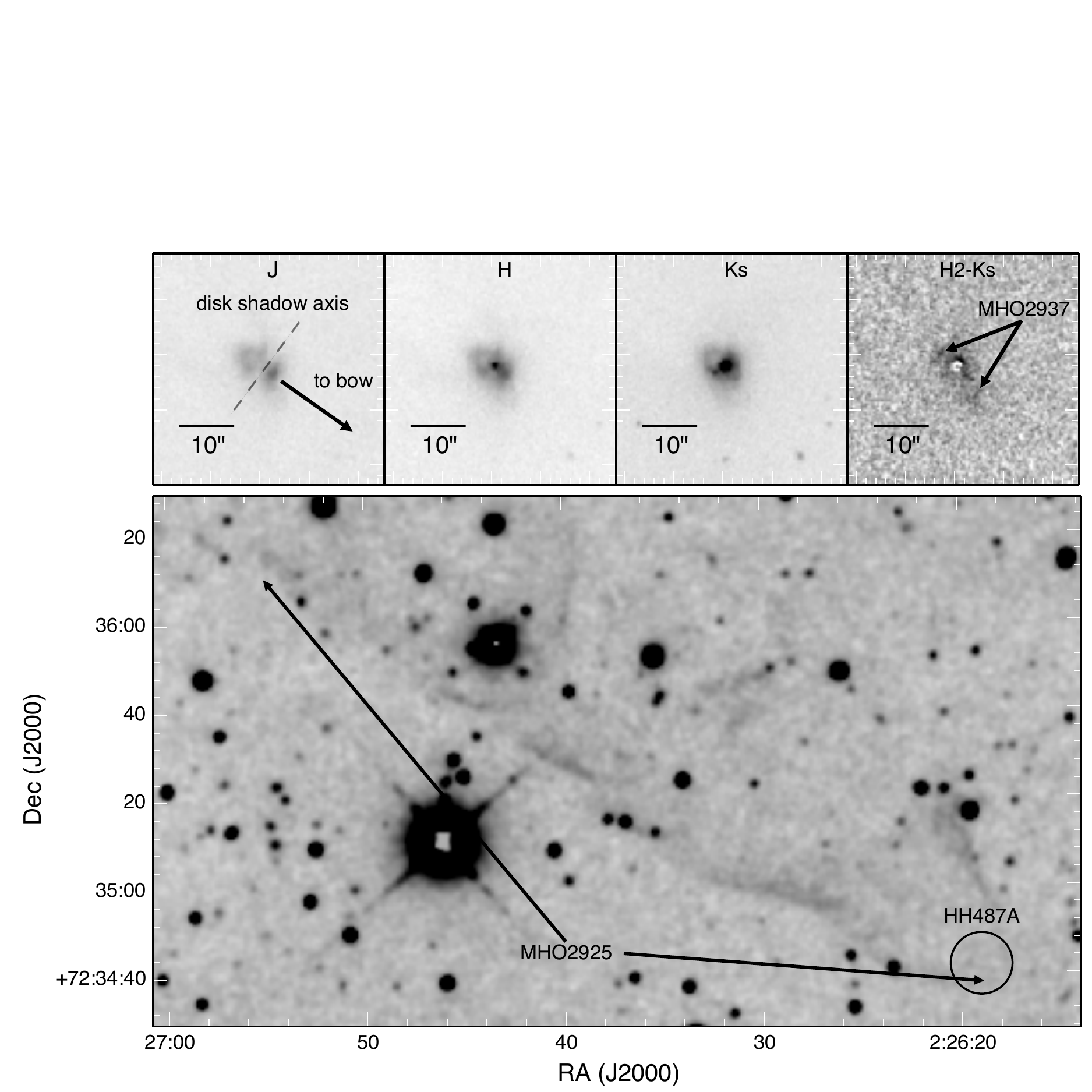}
\caption{An \mh{} image of the MHO\,2925 bow shock discussed in \S\ref{MHO2925}.  The upper panels show the region around SSTSL2\,J022907.88+724347.2 in various filters, including an \mh{}$-$Ks{} difference image which shows the MHO\,2937 jet.}
\label{FigMHO2925}
\end{figure}

MHO\,2925 is a beautiful, filamentary shock system which outlines a bow shock shape (Figs.\ \ref{FigL1340AOverview} \& \ref{FigMHO2925}).  The apex of this bow shock is coincident with HH\,487\,A discovered by \cite{Kumar2002}.  Based on the shock morphology, the outflow source must lie to the northeast somewhere along a line at a position angle of about 54 $\pm$ 1 degrees.

The nearest candidate source along that line is 15.5\arcmin{} away.  That star (SSTSL2\,J022907.88+724347.2) was catalogued by \cite{Kun2016IR} as a Flat SED source.  SSTSL2\,J022907.88+724347.2 is surrounded by a reflection nebula cataloged by \cite{Magakian2003} as their RN3.  Our J, H, and \Ks{} images also show a small reflection nebula surrounding the star (Fig.\ \ref{FigMHO2925}).  In the J image, the reflection nebula is a classic, double lobed shape with an apparent disk shadow obscuring the star itself.  At H-band, the star becomes visible and the double lobed structure of the reflection nebula is less apparent, though a loop-like structure is now visible in the northeast lobe of the nebula.  At \Ks{}, the star is much brighter and the loop-like structure seen in H-band is still visible.

The disk shadow lies along a position angle of 143 $\pm$ 4 degrees, which is perpendicular (within the measurement error) to the position angle defined by the vector from the star to MHO\,2925.

In an \mh{}$-$\Ks{} difference image, there appears to be residual \mh{} emission around the star in a filamentary structure (Fig.\ \ref{FigMHO2925}).  This filament is oriented at position angle 52 $\pm$ 6 degrees and extends from the northeast to the southwest across the star, in reasonable agreement with the direction to MHO\,2925.

Based on the position of the source relative to the perceived flow axis of the MHO\,2925 shock system, the angle of the disk shadow, and the angle of the MHO\,2937 jet, we conclude that the HH\,487\,A, MHO\,2925, and MHO\,2937 shocks are driven by the SSTSL2\,J022907.88+724347.2 source star.  This makes the southwest lobe of this outflow 15.5\arcmin{} in length, which at a distance of 825~pc, corresponds to a length on the sky of 3.7\,pc.

To the northeast of the source, we see some \mh{} shocks about 2.5 arcminutes away near the same axis, but we associate those with the MHO\,2939 flow from IRAS\,F02250+7230 (see \S\ref{MHO2939}), so we are unable to identify any distant components of the northeast lobe of the outflow.

The catalog position of the A3 NH$_3$ core of \cite{Kun2003} lies 28\arcsec{} from the reflection nebula around SSTSL2\,J022907.88+724347.2.  This is in good positional agreement given that the half power beam width and grid size of their map was 40\arcsec{}.

Our association of MHO\,2925 (HH\,487) with SSTSL2\,J022907.88+724347.2 is in disagreement to the conclusion by \cite{Kumar2002} and \cite{Kun2014} that the source of HH\,487 is IRAS\,02224+7227 which lies roughly 6 arcminutes away from the bowshock, but at a position angle of 28.4 degrees.  We believe that the morphology of the \mh{} bow shock is inconsistent with their interpretation.  \cite{Kun2014} discuss the dynamic age of HH~487 if the source were IRAS\,02224+7227, however our association of that shock with SSTSL2\,J022907.88+724347.2 places it 2.5 times as far away and thus the dynamic age of the outflow would be much larger than the 6500 years they estimate.

\subsubsection{A Bright, Curved Outflow:  MHO\,2928, HH\,488\,B, HH\,488\,C}
\label{MHO2928}

\begin{figure}[!htb]
\includegraphics[width=1.0\textwidth]{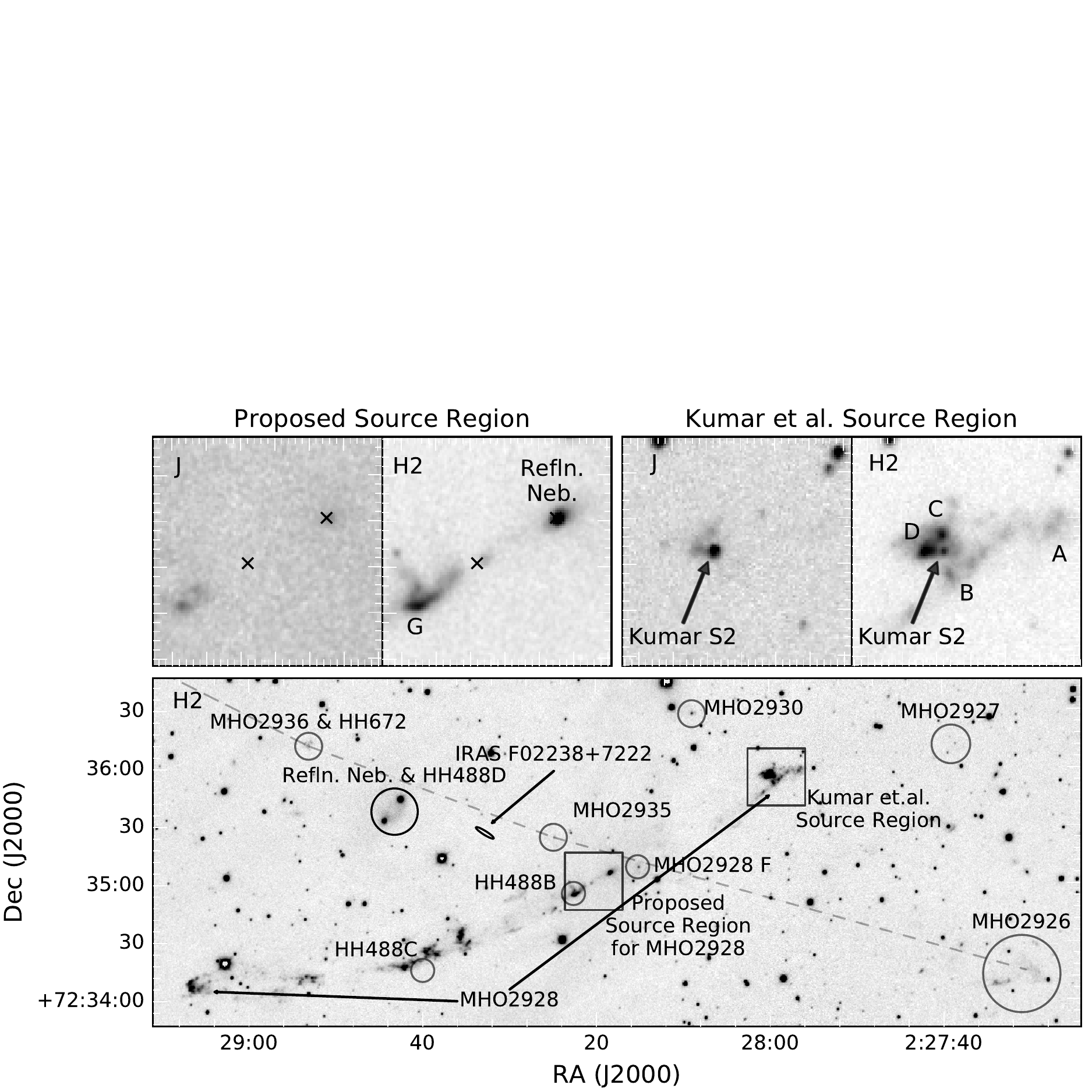}
\caption{J and \mh{} images of the region around MHO\,2928.  The insets along the top show closeup views of our proposed source region for MHO\,2928 and the source region proposed by \cite{Kumar2002}.  In the insets of our proposed source region the positions of SSTSL2\,J022820.81+723500.5 or SSTSL2\,J022820.81+723500.5 are marked with x's and MHO\,2928 knot G and the reflection nebula discussed in the text are labeled.  In the insets of the source region proposed by \cite{Kumar2002}, MHO\,2928 knots A-D are labeled.  \cite{Kumar2002} feature S1 is coincident with knot D.  \cite{Kumar2002} feature J1 is coincident with knot C.  \cite{Kumar2002} feature S2 is labeled.  We find no strong \mh{} feature corresponding to \cite{Kumar2002} feature J2.}
\label{FigMHO2928}
\end{figure}

This flow is a long chain of \mh{} knots and filaments which is about 5.7 arcmin long (Fig.\ \ref{FigMHO2928}), which corresponds to a length of about 1.4\,pc at an assumed distance of 825\,pc.  In this chain of \mh{} knots extending to the southeast, we find knots which correspond to HH\,488\,B and HH\,488\,C of \cite{Kumar2002}.  \cite{Kumar2002} also identified HH\,488\,D which lies $\sim$1.3\arcmin{} North of the chain of \mh{} knots.  We believe this shock is not associated with this flow and it is discussed elsewhere in this paper (see \S\ref{MHO2926}).

The gradual curve of the shock system suggests that either the source is moving to the southwest or that the source star is precessing and thus changing the launch angle of the flow.

\cite{Kumar2002} suggested that the flow was launched from a star which corresponds to SSTSL2\,J022759.92+723556.4, a probable double star.  \cite{Kun2016IR} catalog it as a flat SED source, but note that the SED "results from the composite fluxes of the central objects".  This region is very complex and \cite{Kumar2002} identified two candidate jets in \Ha{} and \sii{} emerging from two stars (which they designated S1 and S2) separated by about 2 arcseconds at position angles of 320 and 275 degrees respectively.  S1 is the brighter of the two in \Ha{}, but S2 is the only one visible in their continuum (Gunn z) image.

In our broadband J (see Fig.\ \ref{FigMHO2928}) and H (not shown) images there are 5 objects within 8 arcseconds of the Kumar source.  Two of these are 2.7 arcsec to the East and 2.0 arcsec to the North respectively, there appears to be extended emission connecting these two points to the source star.  The arrangement of these three sources appears to match the arrangement of \cite{Kumar2002} sources S1 and S2 and J1 jet, however, the positions of S1 and S2 listed in \cite{Kumar2002} lie about 1.5 arcsec North of the corresponding objects in our images.  This may be due to an offset in the WCS in the \cite{Kumar2002} coordinates.  For this discussion we will assume that the \cite{Kumar2002} sources correspond to the objects visible in our images.  In this case, S2 of \cite{Kumar2002} is clearly a stellar component in our images.

Whether the other two objects in our images are stars, reflection nebulae, or shocks is not completely clear.  The source corresponding to \cite{Kumar2002} S1, however, is very bright in \mh{}, so we suspect that it is shocked gas rather than a stellar source, so we consider it an \mh{} shock (MHO\,2928\,D).  The object to the northwest (corresponding to \citealt{Kumar2002} J1) is also very bright in \mh{} so we similarly consider it an \mh{} shock (MHO\,2928\,C).  We see no object corresponding to \cite{Kumar2002} object J2 which lies ~3 arcsec due West of \cite{Kumar2002} S2.

In the \mh{} filter, a filamentary chain of knots (MHO\,2928\,B) is visible about 4 arcsec southwest of the Kumar source (Fig.\ \ref{FigMHO2928}).  The continuous part of the filament is about 23 arcseconds long along a position angle of 139 degrees.  This filament does not pass through the Kumar source star, so it is unlikely that this is a jet, but is possibly the shocked wall of the outflow cavity.  It should be noted that this filament does not correspond to the jet features in \cite{Kumar2002} as it lies south of the source star while the \cite{Kumar2002} features are coincident with the star.

After inspection of the large scale structure of the outflow and of the WISE 3.4\um{}, 4.6\um{}, 12\um{}, \& 22\um{} images we conclude that the source of the flow is not in the region suggested by \cite{Kumar2002}, but that it lies 1.6-1.8\arcmin{} to the southeast along the flow axis and that the flow is driven by either SSTSL2\,J022820.81+723500.5 or SSTSL2\,J022818.51+723506.2.  We prefer this as the source region (similar to the conclusion of \citealt{Kun2016IR}) instead of that suggested by \cite{Kumar2002} because these two candidate source stars were both classified as Class 0/I by \cite{Kun2016IR} and SSTSL2\,J022820.81+723500.5 was discussed as a candidate Class 0 source.  In addition, the filamentary \mh{} structure around these two sources (MHO\,2928\,G) passes through the sources themselves and could plausibly be a jet, unlike the filament (MHO\,2928\,B) south of the \cite{Kumar2002} source star.  Lastly, these two sources are coincident with the A1 NH$_3$ core of \cite{Kun2003}.  Of the two sources, SSTSL2\,J022818.51+723506.2 is coincident with a reflection nebula visible in our J, H, and \Ks{} images (see inset in Fig.\ \ref{FigMHO2928}) while the SSTSL2\,J022820.81+723500.5 is invisible in J and H, and only faintly visible in \Ks{} and \mh{}.

\subsubsection{A Candidate Large Scale Outflow:  MHO\,2926, MHO\,2935, MHO\,2936, MHO\,2938, MHO\,2940, HH\,488\,D, HH\,672}
\label{MHO2926}

MHO\,2926 (Fig.\ \ref{FigMHO2928}) is a complex of knots which makes up a nice bow shock pointing roughly back to a source which must lie along PA $\sim$ 60-80 degrees.  Along that line lie HH\,488\,D and HH\,672 in addition to several \mh{} shocks (MHO\,2928, 2935, 2936, 2938, 2940).

MHO\,2935 (Fig.\ \ref{FigMHO2928}) is a compact \mh{} knot.  HH\,488\,D has no clear \mh{} counterpart, but is coincident with reflection nebulosity visible in our J, H, and \Ks{} images.  HH\,672 has an \mh{} counterpart (MHO\,2936; Fig.\ \ref{FigMHO2928}) which is a small cluster of \mh{} knots.  MHO\,2938 (Fig. \ref{FigL1340AOverview}) is a pair of small knots separated by about 12\arcsec{}.

While the association of these features is not conclusive, we find it likely that there is a flow along PA $\sim$75 degrees, emanating from one of the four candidate sources described below.  This flow would be comprised of MHO\,2926, MHO\,2935, HH\,488\,D, MHO\,2936 (and its optical counterpart HH\,672), and MHO\,2938 (Fig.\ \ref{FigL1340AOverview}) with the likelihood that some of the \mh{} emission in MHO\,2928 is also part of this flow where the two flows cross.  If this is the case, this flow is 9.9\arcmin{} long which would correspond to 2.4\,pc.  In addition, this flow has the shape of a gentle arc similar to the MHO\,2928 flow.  The opening of the arc faces back toward RNO7.  If the arc is due to the source being in motion, then this star could potentially have been dynamically ejected from the cluster core.

The knot MHO\,2940 (Fig. \ref{FigL1340AOverview}) lies another 3.4 arcminutes northeast of MHO\,2938 along the flow axis.  It may be a distant shock in this flow, but the association is unclear because it also lies near the axis of the MHO\,2939 flow.

There are several candidate sources which lie near the axis of this flow.  From east to west, they are:  SSTSL2\,J022844.40+723533.5, SSTSL2\,J022842.57+723544.3, IRAS\,F02238+7222, and WISE\,J022817.97+723517.5.  We briefly discuss each in the paragraphs below.

The reflection nebula coincident with HH\,488\,D (Fig.\ \ref{FigMHO2928}) spans a region between two stars.  The southern star is SSTSL2\,J022844.40+723533.5 and the northern is SSTSL2\,J022842.57+723544.3.  Both were classified by \cite{Kun2016IR} as Class 0/I sourced based on their SED slopes.

IRAS\,F02238+7222 (identified by \citealt{Kumar2002} as a candidate young star) also lies along the axis of the flow, however we find no sources in the WISE images or in our near-IR images which seem to correspond to IRAS\,F02238+7222.  We conclude that this is likely a background extragalactic source.

Lastly, there is WISE\,J022817.97+723517.5, a star near the intersection of the MHO\,2928 flow and this flow's axis.  This star was not selected as a protostar in \cite{Kun2016IR} and lies somewhat outside the color parameters which \cite{KoenigLeisawitz2014} use to classify WISE sources as protostars, however, it is visible in our \Ks{} image and appears to have a faint filament of \mh{} emission (MHO\,2928\,F) which passes through the star along roughly the same axis that the flow is expected to occupy.  Because this filament overlaps with MHO\,2928 (and is catalogued as part of that MHO shock complex), it is impossible to tell if it is part of this flow or not, but the morphology is suggestive.  WISE\,J022817.97+723517.5 also lies within the contours of the the A1 NH$_3$ core of \cite{Kun2003}.  Although this core likely contains several stars as the source of the MHO\,2928 flow described in \S\ref{MHO2928} is also present in this core.  Because of the confused nature of the emission in this region (much of it presumably from the MHO\,2928 flow), the coincidence of the WISE star and a shock filament is not as compelling as it would be in other regions, therefore we favor one of the two SSTSL sources mentioned above as the source for this flow.

\subsubsection{MHO\,2932, HH\,671\,A}
\label{MHO2932}

\cite{Magakian2003} found two HH knots in the RNO7 region (HH\,671\,A and B) which are separated by about 1 arcminute.  Both objects have bright, compact \mh{} counterparts.  

\begin{figure}[!htb]
\includegraphics[width=0.5\textwidth]{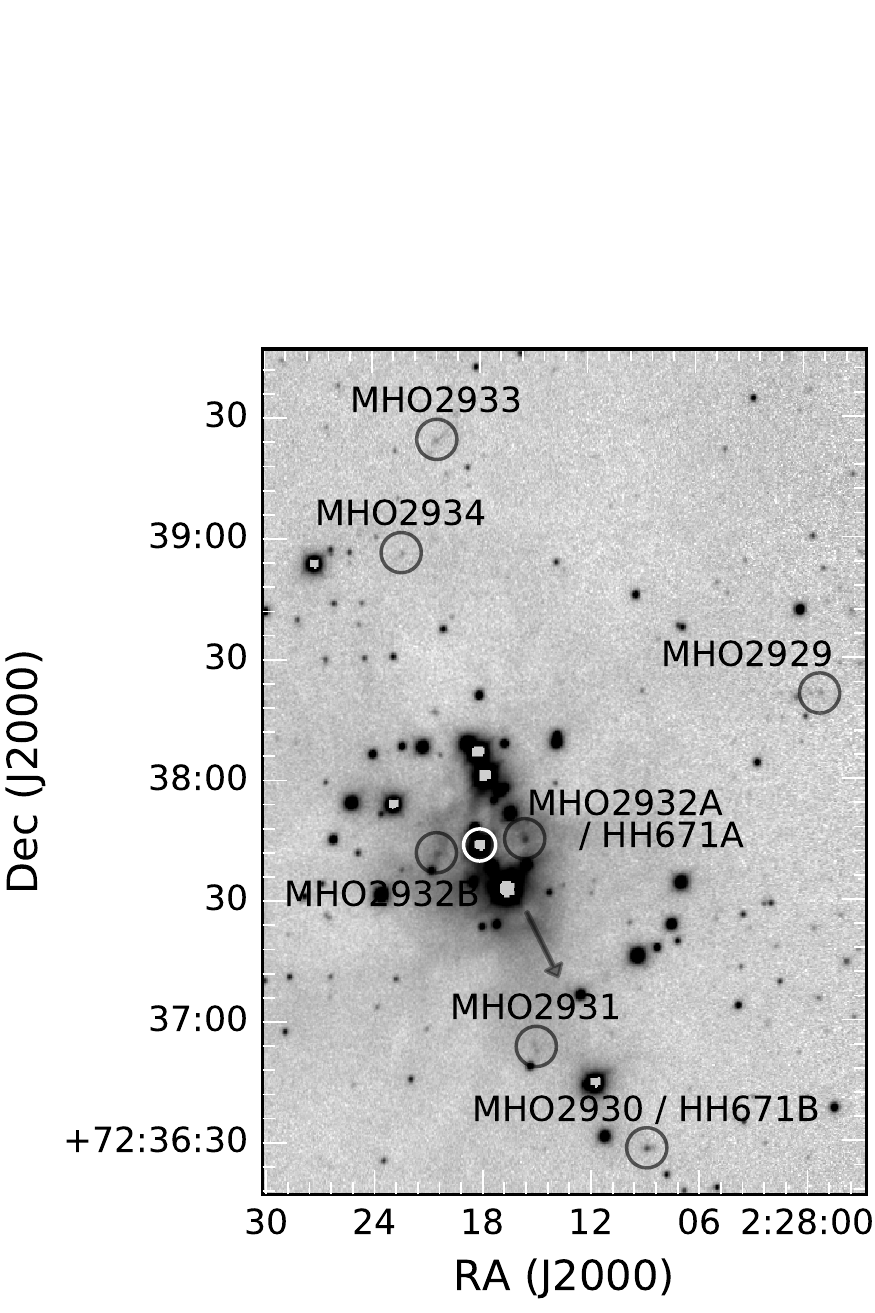}
\caption{An \mh{} image of the region around MHO\,2932 discussed in \S\ref{MHO2932}.  The white circle indicates the position of WISE\,J022818.07+723743.8 (\cite{Magakian2003} \ha{} star number 8).  The gray arrow indicates the approximate position and orientation of the jet discussed in \cite{Kumar2002} \S3.1.}
\label{FigMHO2932}
\end{figure}

The counterpart to HH\,671\,A (MHO\,2932; Fig.\ \ref{FigMHO2932}) lies 11.2\arcsec{} West of one of the brighter stars in the RNO7 cluster.  There is an additional knot (MHO\,2932~B) which lies 10.6\arcsec{} East of the same star.  Their positions opposite one another across the star suggest that they may be the east and west components of an outflow which is launched from that star at a position angle of 99$\pm$2 degrees.  The source star was cataloged as an \ha{} emission line star by both \cite{Magakian2003} (who designated it \ha{} star number 8) and by \cite{Kun2016Opt} (star 19 in their Table\,2).

The association of these shocks with a particular source star is tenuous at best due to the crowded field (several other \ha{} emission line stars lie in the RNO7 cluster) and to the lack of resolved morphology in the MHO objects.  It should be noted this is in contradiction to the connection by \cite{Magakian2003} of HH\,671\,A and HH\,671\,B into a single flow which is implied by giving them both the same HH number (HH\,671) and distinguishing them by using the knot A and knot B designations.

\subsubsection{Additional Shocks Near the RNO7 Cluster: MHO\,2929, MHO\,2930, MHO\,2933, MHO\,2933, MHO\,2934}
\label{RNO7Region}

Five additional shocks lie within $\sim$1.6\arcmin{} of the RNO7 cluster (Fig.\ \ref{FigMHO2932}).  To the south lie MHO\,2930 and MHO\,2931.  MHO\,2930 is the counterpart to HH\,671\,B.  It is a bright, compact knot of \mh{} emission, which lies about 1.5\arcmin{} south of the center of the RNO7 cluster.  MHO\,2931 is a fainter, more diffuse feature which lies closer to the cluster (1.0\arcmin{} south of RNO7).

To the North of RNO7 lie MHO\,2934 and MHO\,2933 at distances of 1.2\arcmin{} and 1.6\arcmin{} respectively.  MHO\,2934 is a small faint, compact knot, while MHO\,2933 is a 12\arcsec{} long filament, oriented roughly NW-SE.

These four objects may make up the northern and southern components of one or more flows emerging from one of the stars in the RNO7 cluster, however, clear association with a particular flow or source star is not possible with these images.

MHO\,2929, which lies 1.5\arcmin{} West of RNO7, consists of a single compact \mh{} knot.

\cite{Kumar2002} describe a candidate HH jet in this region.  Their Figure 2b labels it as emerging from an \ha{} emission line star (labeled Ha1 in their Figure 2c).  The feature is roughly 10 arcseconds long, is stronger in \sii{} than \ha{}, and is oriented to the southwest.  It points generally in the direction of MHO\,2930, some bright knots in MHO\,2928, MHO\,2927, and MHO\,2926.  However, we do not associate it with any of those as the alignments are not convincing and we can not confirm the existence of this jet as it does not show up in our near-IR images (Fig.\ \ref{FigMHO2932}).

\subsubsection{MHO\,2927}

MHO\,2927 (Fig.\ \ref{FigMHO2928}) lies about 3.7 arcminutes southwest of RNO7.  It has the appearance of a filament, about 9 arcsec long oriented along PA $\sim$60 degrees with a brighter, possibly bow-shaped structure on the west end.  A line drawn along PA $\sim$60 degrees extending to the northeast away from the bowshock-like structure passes near the southern edge of the RNO7 cluster fo stars, suggesting that MHO\,2927 may be launched by a source in that region, however the shock is faint making the morphological link of this with any particular source tenuous at best.  In addition, three nearby flat spectrum SED protostars identified by \cite{Kun2016IR} lie closer to this shock than RNO7 at position angles ranging from 75 to 85.

\subsubsection{MHO\,2939, HH\,489}
\label{MHO2939}

\begin{figure}[!htb]
\includegraphics[width=0.5\textwidth]{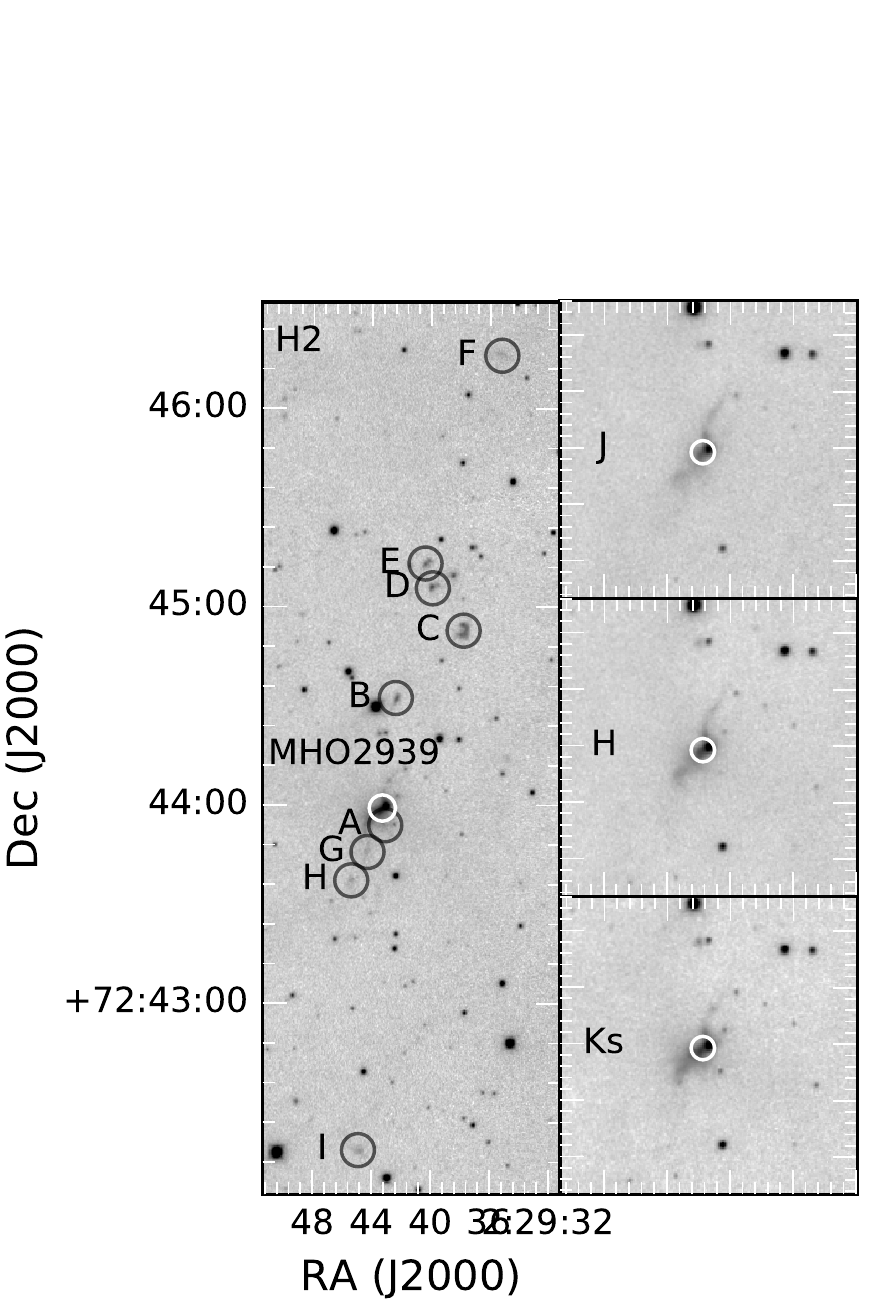}
\caption{A set of \mh{}, J, H, \& \Ks{} images of the region around MHO\,2939 discussed in \S\ref{MHO2939}.  A white circle indicates the position of the binary source discussed in the text.}
\label{FigMHO2939}
\end{figure}

\cite{Kumar2002} found two knots (HH\,489\,A and B) which lie roughly symmetrically 1 arcminute on either side of an IRAS source.  \cite{Kumar2002} also found a "wisp of nebulosity seen in Ha emission [which] extends out from both sides of the source along the flow direction."  In our continuum (J, H, \Ks{}) images (Fig.\ \ref{FigMHO2939}), a filamentary reflection nebula is visible around the source along a similar position angle to the filament described by \cite{Kumar2002}.  

The IRAS source corresponds to a pair of Class\,I sources identified by \cite{Kun2016IR} as SSTSL2\,J022943.01+724359.6 and SSTSL2\,J022943.64+724358.6 which are separated by 2.8\arcsec{}.  They lie within the A4 NH$_3$ core of \cite{Kun2003}.  One of these sources is presumably the driving source of the HH\,489 flow.

Our \mh{} image (Fig.\ \ref{FigMHO2939}) reveals several shocks in the region (designated MHO\,2939) distributed along a line which corresponds to the position angle of the filament described by \cite{Kumar2002} and by the filament visible in the J, H, and \Ks{} images.  Our shocks trace an outflow which is 4 arcminutes long at a position angle of $\sim$170 degrees.

The knot MHO\,2940 lies not far off of this axis about 4.5 arcminutes South of the source along a position angle of 157\arcdeg{}.  However, it also lies along the axis of the MHO\,2926 flow, so it is not clear which flow it is associated with.

\subsection{L1340B}\label{L1340B}

\subsubsection{A Large S-Shaped Outflow:  MHO\,2942}
\label{MHO2942}

\begin{figure}[!htb]
\includegraphics[width=1.0\textwidth]{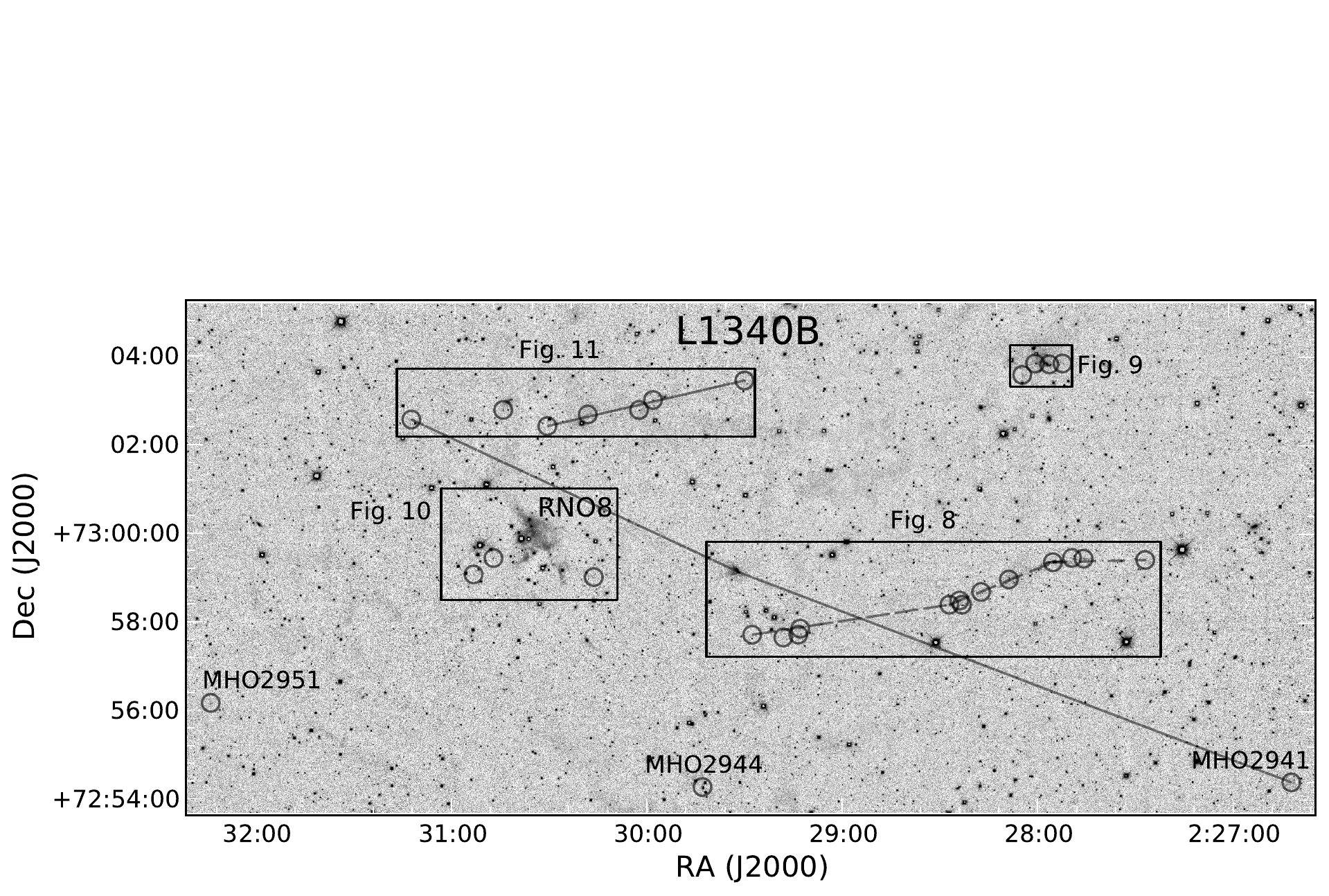}
\caption{An \mh{} image of the L1340\,B region.  Circles mark the positions of \mh{} shocks and boxes indicate the locations of subsequent figures in the text.}
\label{FigL1340BOverview}
\end{figure}

\begin{figure}[!htb]
\includegraphics[width=1.0\textwidth]{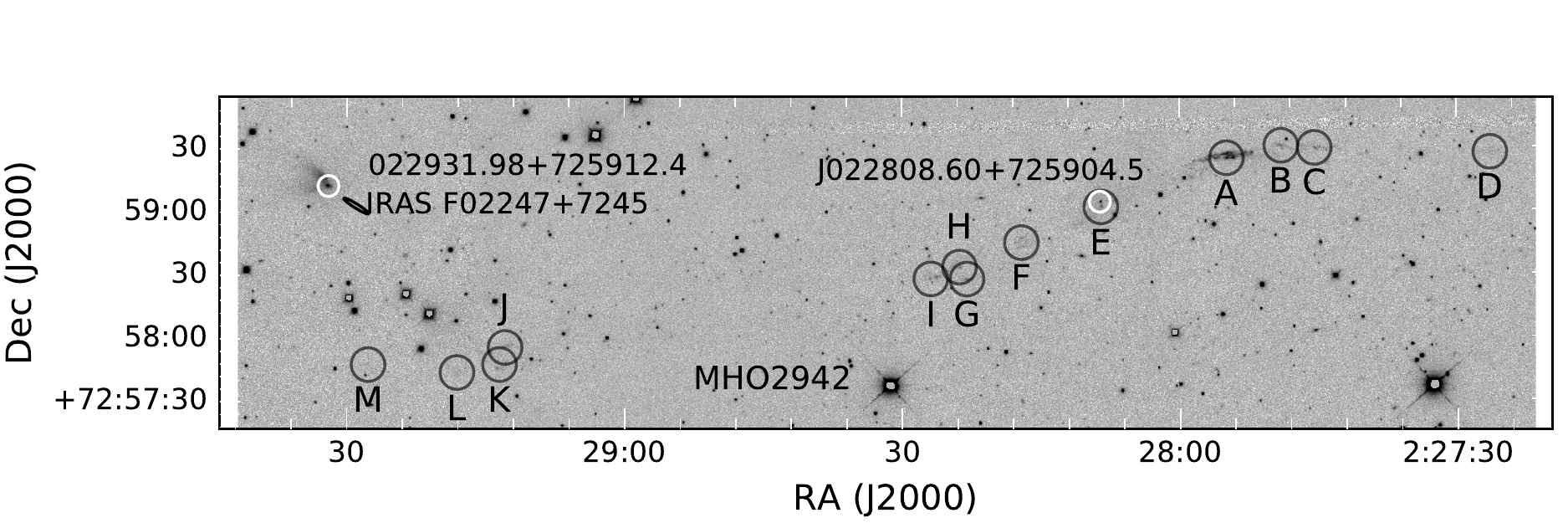}
\caption{An \mh{} image of the MHO\,2942 S-shaped outflow discussed in \S\ref{MHO2942}.  The V-shaped reflection nebula around IRAS\,F02247+7245 (discussed in \S\ref{IRASF02247+7245}) is visible at the upper left corner of the image.  The error ellipse for IRAS\,F02247+7245 is also marked in black.  The SSTSL2 sources discussed in the text are marked with white circles and labeled.}
\label{FigMHO2942}
\end{figure}

MHO\,2942 (Figs.\ \ref{FigL1340BOverview} \& \ref{FigMHO2942}) is a 9.1\arcmin{} long chain of \mh{} knots orignally discovered in our 2005/2006 NICFPS run at Apache Point.  The chain is oriented roughly East-West with a gentle S-shaped curve.  \cite{Kun2016IR} found a candidate Class\,0 source (SSTSL2\,J022808.60+725904.5) near this flow axis: 2.6\arcsec{} North of the brightest condensation in knot E.

\subsubsection{The IRAS\,F02247+7245 Outflow: MHO\,2941}
\label{IRASF02247+7245}

About 1.4\arcmin{} North of the easternmost knots in MHO\,2942 lies a dramatic V-shaped reflection nebula (visible in \mh{} in the upper left corner of Fig.\ \ref{FigMHO2942}) with a star at its apex.  The reflection nebula is visible at J, H, and \Ks{}, but the star (SSTSL2\,J022931.98+725912.4) is only apparent in the \Ks{} image.  IRAS\,F02247+7245 lies 15 arcsec away from the star at the apex of the reflection nebula on a position angle of 54 degrees.  The uncertainty ellipse of the IRAS source is 14 arcseconds along the major axis along a position angle of 58 degrees.  This places the \cite{Kun2016IR} source just outside of the error ellipse, but we believe that the IRAS source corresponds to the \cite{Kun2016IR} source which they identified as a candidate Class\,0 source.

The V-shaped reflection nebula opens to the East and is bright on the north and south edges and dark along its central axis which lies along a position angle of roughly 70 degrees.  Roughly 13.3\arcmin{} to the southwest, opposite the opening of the reflection nebula (along a position angle of 249 degrees), lies MHO\,2941, which is a faint bow-shaped arc of \mh{} emission.  Based on the agreement in position angle with the opening of the reflection nebula, we find it likely that MHO\,2941 is a distant bow shock in a flow driven by IRAS\,F02247+7245.  This means that one lobe of this flow is 13.3\arcmin{} (3.2\,pc) long.

To the northeast, 8.1 arcminutes away along a position angle of 65 degrees, lies MHO\,2948.  The alignment of this shock relative to the axis of the reflection nebula makes the association of this shock to the IRAS\,F02247+7245 source compelling.  On the other hand, this shock lies near the axis of, and may be part of, the MHO\,2946 flow (see Fig.\ \ref{FigL1340BOverview} \& \ref{FigMHO2946}).

\cite{Kun2016IR} also describe a "jet-like feature, bright at 4.5\,\um{} on the western side (Fig.\,20)."  By inspection of their Fig.\,20, we see that this jet-like feature emerges at a position angle of roughly 275 degrees.  This does not align with a line drawn to MHO\,2948 and we find no shocked emission in that direction other than MHO\,2942 which is clearly associated with a different source star.

We hypothesize that the jet-like feature in the Spitzer image is not a jet, but is the wall of the outflow cavity.  While the eastern cavity walls are visible in our J, H, and \Ks{} images (see Fig.\ \ref{FigMHO2942}), there is no corresponding western cavity.  We suggest that this is due to greater extinction on that half of the outflow, either due to a local feature in the cloud structure or due to the ouflow being aimed toward Earth (i.e. we predict it is a blueshifted outflow lobe) while the western lobe is more highly extincted.  Indeed, \cite{Kun2016IR} note the "color difference between the eastern and western nebulosities" which in their Fig.\,20 can be seen as this feature being brighter in green (4.5\,\micron{}) than in the blue (3.6\,\micron{}).  This could easily be due to a difference in extinction along the line of sight to the two cavity walls.

\subsubsection{A Compact S-Shaped Outflow:  MHO\,2943}
\label{MHO2943}

\begin{figure}[!htb]
\includegraphics[width=0.5\textwidth]{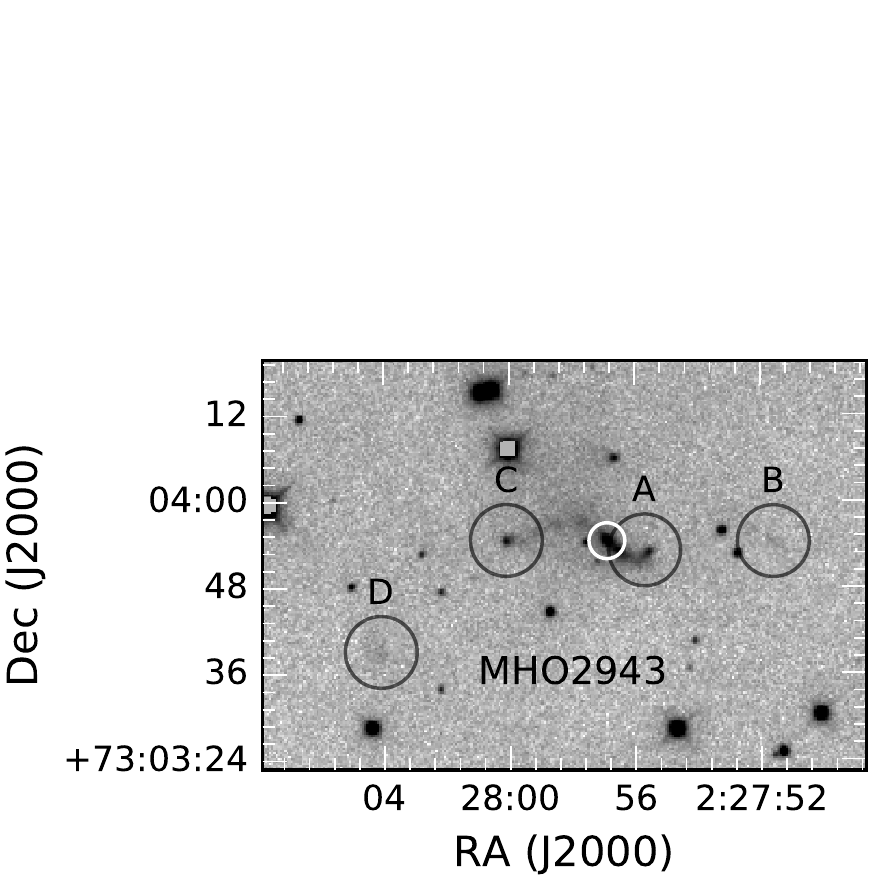}
\caption{An \mh{} image of the MHO\,2943 flow discussed in \S\ref{MHO2943}.  The source star (SSTSL2\,J022756.91+730354.4) lies in the brightest portion of the reflection nebula.  The \mh{} shocks are marked and the position of the WISE source discussed in the text is indicated with a white circle..}
\label{FigMHO2943}
\end{figure}

MHO\,2943 (Fig.\ \ref{FigMHO2943}) is a set of four \mh{} knots which appears to emerge from the region surrounding SSTSL2\,J022756.91+730354.4 which \cite{Kun2016IR} identified as a Class\,0/I source.  The star is visible in our \Ks{} and H images, but disappears in J.  It is surrounded by a filamentary S-shaped reflection nebula about 20 arcseconds across.  

Two of the four MHO\,2943 knots are very compact and bright and lie nearest the source.  One 6.4 arcseconds to the West, the other 14 arcseconds to the east.  The other two knots are fainter, more diffuse and lie 24 arcseconds to the West and 36 arcseconds to the east respectively.

\subsubsection{The RNO8 Region:  MHO\,2945, MHO\,2949, MHO\,2950}
\label{RNO8Region}

\begin{figure}[!htb]
\includegraphics[width=0.5\textwidth]{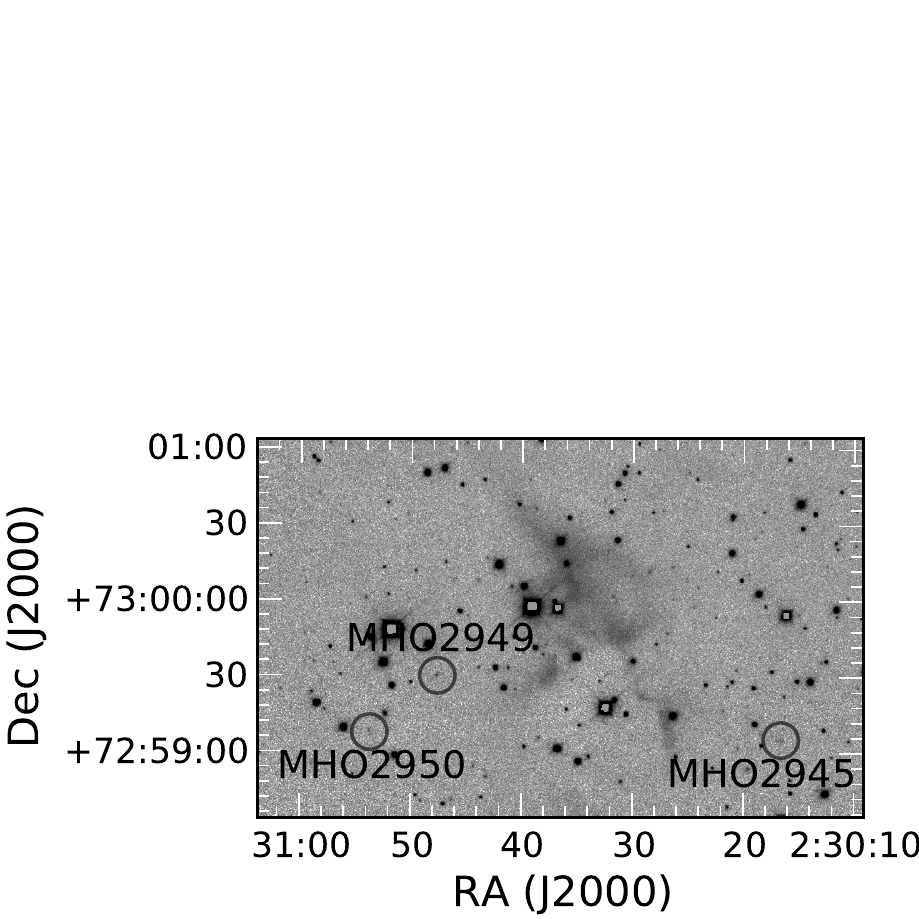}
\caption{An \mh{} image of the RNO8 region discussed in \S\ref{RNO8Region} showing the locations of MHO\,2945, MHO\,2949, \& MHO\,2950.}
\label{FigMHO2945}
\end{figure}

Three \mh{} knots lie in a 1.4 arcminute diameter region roughly centered on southern half of the RNO8 reflection nebula.  All three knots are relatively faint.  MHO\,2945 and MHO\,2950 are diffuse, while MHO\,2949 is a compact knot (Fig.\ \ref{FigMHO2945}).  The MHO\,2945 and MHO\,2949 knots lie nearly equidistant (70\arcsec{} and 67\arcsec{} respectively) on either side of SSTSL2\,J023032.44+725918.0, suggesting that they may be launched from this star, however, this may be a chance alignment.  \cite{Kun2016IR} classified SSTSL2\,J023032.44+725918.0 as a Class\,0/I source, however they also describe its classification as ambiguous and declare that it flass "into the Class II regime near the Class I/ Class II boundary".  \cite{Kun2016Opt} found it to be a "late G spectral type with the Balmer lines in emission" (star 43 in their Table\,2) and argue that it is a star surrounded by a disk seen at high inclination (leading to the relatively low extinction to the stellar photosphere).

\subsubsection{MHO\,2946}
\label{MHO2946}

\begin{figure}[!htb]
\includegraphics[width=1.0\textwidth]{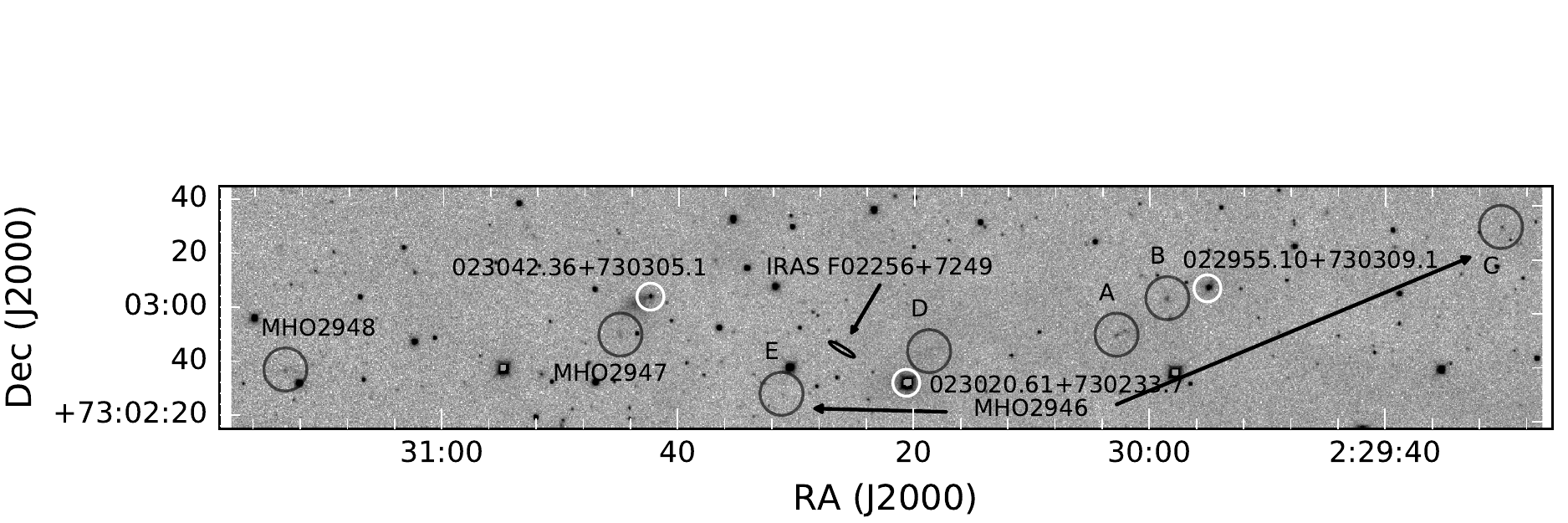}
\caption{An \mh{} image of the MHO\,2946 flow discussed in \S\ref{MHO2946}.  The position of IRAS\,F02256+7249 is indicated with an ellipse which corresponds to the positional error of the source in the IRAS catalog.  The reflection nebula surrounding SSTSL2\,J023042.36+730305.1 is visible in the left half of the image.  The positions of SSTSL2\,J023020.61+730233.7 and SSTSL2\,J022955.10+730309.1 are indicated with white circles.}
\label{FigMHO2946}
\end{figure}

MHO\,2946 (Fig.\ \ref{FigMHO2946}) is a chain of \mh{} knots about 4.5 arcminutes long.  The axis defined by the knots passes near three candidate source stars.  

The western of these candidates is SSTSL2\,J023042.36+730305.1, a Class\,0/I star identified by \cite{Kun2016IR}.  This star has a small, faint wisp of \Ks{} emission which may be a reflection nebula emerging on the western side of the star (see also \S\ref{MHO2947}).  This source star lies furthest from the apparent axis of MHO\,2946, so we do not favor it as a source candidate.

Another candidate source is SSTSL2\,J023020.61+730233.7, which \cite{Kun2016IR} cataloged as a flat spectrum source.  This star is bright at all near-IR wavelengths (J, H, \Ks{}) and is even visible faintly in the POSS R image, though not the POSS B image.  IRAS\,F02256+7249 lies about 18 arcseconds from this second candidate star.  The eastern candidate source star is SSTSL2\,J022955.10+730309.1, another Class\,0/I star in the \cite{Kun2016IR} catalog.  The B1 NH$_3$ core of \cite{Kun2003} lies roughly in between these two more eastern sources.  

\subsubsection{MHO\,2947}
\label{MHO2947}

Roughly 40 arcseconds northeast of the easternmost component of MHO\,2946 lies a fan shaped reflection nebula with a star (SSTSL2\,J023042.36+730305.1) at its apex (see also \S\ref{MHO2946}).  The B2 NH$_3$ core of \cite{Kun2003} lies 0.5\arcmin{} northwest of the apex of the reflection nebula. The nebula opens toward position angle $\sim$120 and is visible in all of our J, H, \& \Ks{} images, while the star is invisible at J and H, but bright in \Ks{}.  About 18 arcseconds southeast of the star along PA $\sim$140 degrees lies a faint, diffuse \mh{} knot (MHO\,2947) which is likely driven by the source embedded in that reflection nebula.  It is also possible, however, that this is a distant shock in the MHO\,2946 flow.

\subsubsection{Distant \mh{} Shocks:  MHO\,2951, MHO\,2944}

Two shocks in the L1340\,B region lie far from the other groupings.  MHO\,2944 (Fig.\ \ref{FigL1340BOverview}) lies in the southern part of this region.  It is a bright, compact \mh{} knot which lies roughly equidistant from two Class\,0/I sources identified by \cite{Kun2016IR}: SSTSL2\,J022932.31+725503.2 (also \ha{} emission line star number 38 in \cite{Kun2016Opt} Table\,2) and SSTSL2\,J022949.62+725326.1.

MHO\,2951 is a faint, diffuse knot (Fig.\ \ref{FigL1340BOverview}) which lies about 3 arcminutes southeast of four sources identified as either Class\,0/I or Flat SED by \cite{Kun2016IR}.

\subsection{L1340C}\label{L1340C}

\subsubsection{The RNO9 Region}
\label{RNO9region}

\begin{figure}[!htb]
\includegraphics[width=1.0\textwidth]{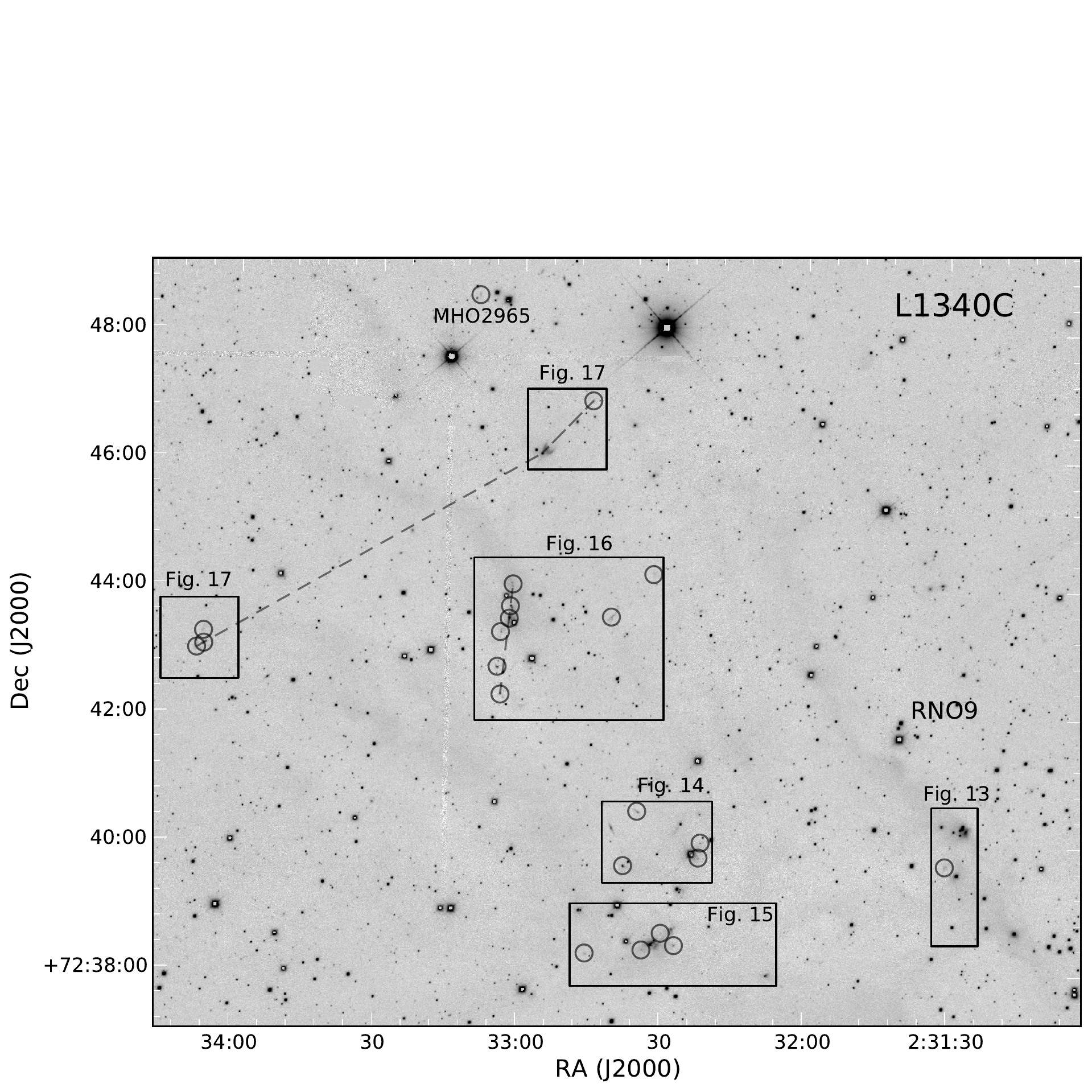}
\caption{An \mh{} image of the L1340C region.  Circles mark the positions of \mh{} shocks and boxes indicate the locations of subsequent figures in the text.}
\label{FigL1340COverview}
\end{figure}

\begin{figure}[!htb]
\includegraphics[width=0.5\textwidth]{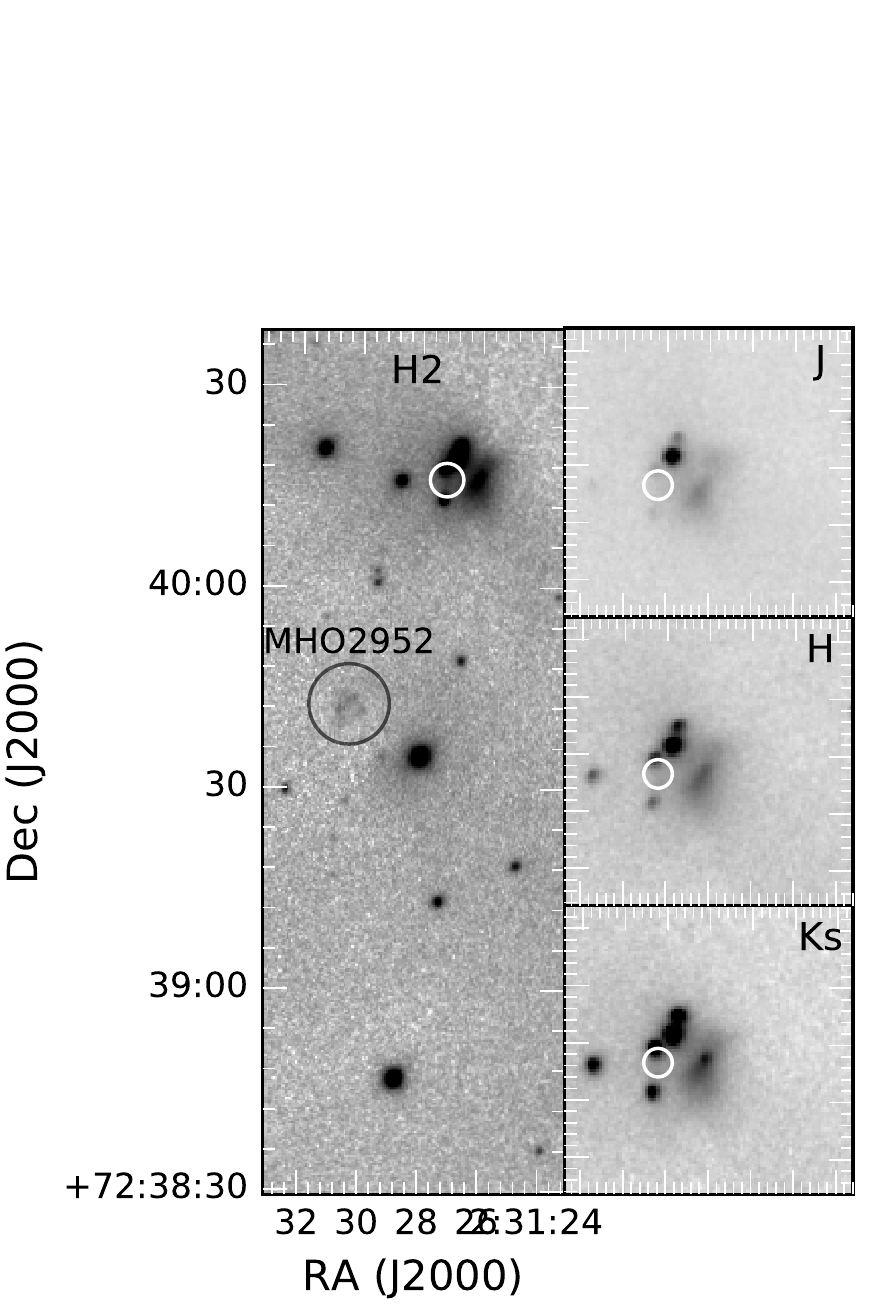}
\caption{The left panel shows an \mh{} image of the MHO\,2952 flow discussed in \S\ref{RNO9region}.  The right hand panels show broadband images of the source region with the reflection nebula discussed in the text.  The white circle indicates the catalog position of the SSTSL source discussed in the text.}
\label{FigMHO2952}
\end{figure}

The RNO9 region contains a grouping of stars roughly coincident with "a nebulous star that is bright at 2 um and invisible in the optical wavelengths" described by \cite{Kumar2002}.  In this grouping lies SSTSL2\,J023127.34+724012.9 which \cite{Kun2016IR} classified as a Class\,0/I source.  Just to the west of that source is a star visible in our \Ks{} image which has a small bipolar reflection nebula oriented at about PA $\sim$150 degrees around it (Fig.\ \ref{FigMHO2952}).  The star is invisible in our J image, but bright at H and \Ks{}.  The reflection nebula appears to have a disk shadow roughly perpendicular to the bipolar axis of the nebula in the J image which disappears at H and \Ks{}.  We find a \mh{} knot (MHO\,2952) 39 arcsec southeast of this object at PA $\sim$156 degrees.  Thus, we find it likely that the star with the small bipolar reflection nebula is driving an outflow along that axis.

\subsubsection{The L1340\,C Mid-IR Cluster North:  MHO\,2953, MHO\,2954, MHO\,2959, MHO\,2960}
\label{MHO2953}

Roughly 4.3 arcminutes southeast of RNO9 is a grouping of three IRAS sources (F02277+7226, 02276+7225, F02279+7225) listed by \cite{Kun1994} as candidate young stars.  The WISE 22\um{} image of this region reveals eight stars in a 2.7\arcmin{} diameter region.  This cluster of infrared sources stands out from other regions of the L1340 cloud complex as being particularly dense.  The only comparable cluster in the L1340 region is RNO7 in L1340\,A.

There is a group of four \mh{} knots (MHO\,2953, 2954, 2959, 2960) in the northern half of this cluster near IRAS\,F02277+7226.  This is also roughly coincident with the C2 NH$_3$ core of \cite{Kun2003}.  Two Class\,0/I sournces from \cite{Kun2016IR} lie in this region: SSTSL2\,J023225.98+724020.1 and SSTSL2\,J023237.90+723940.7. 

\begin{figure}[!htb]
\includegraphics[width=0.5\textwidth]{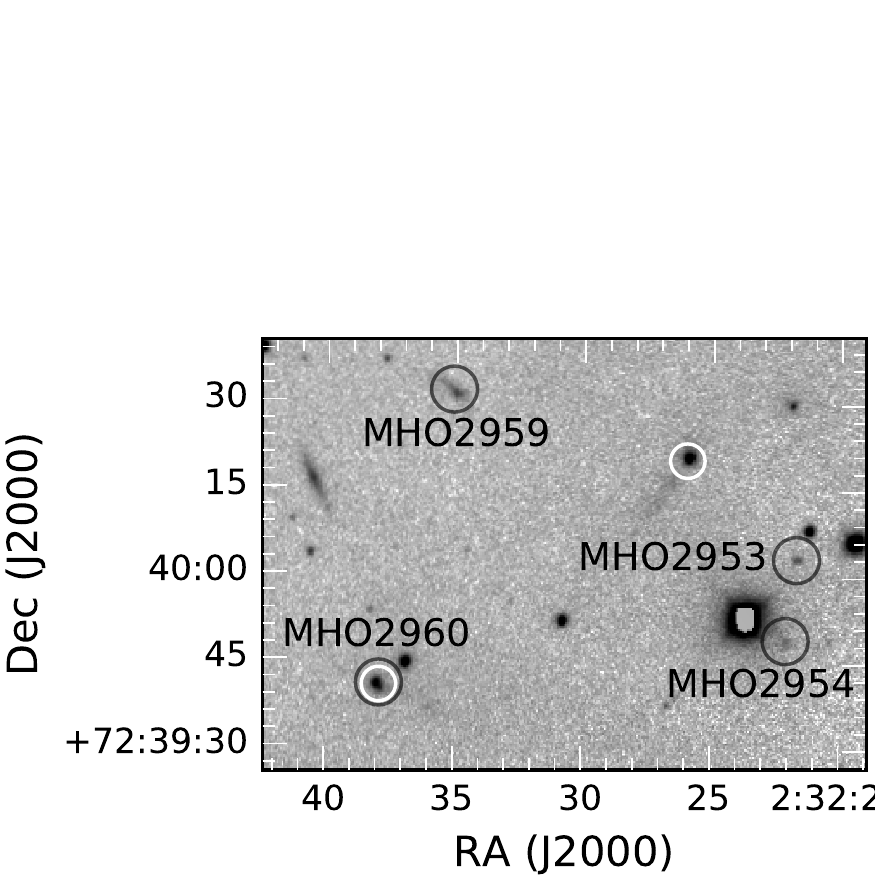}
\caption{An \mh{} image of the region discussed in \S\ref{MHO2953} containing MHO\,2953, MHO\,2954, MHO\,2959, \& MHO\,2960.  The white circles represent the positions of SSTSL2\,J023225.98+724020.1 and SSTSL2\,J023237.90+723940.7.}
\label{FigMHO2953}
\end{figure}

MHO\,2960 is coincident with SSTSL2\,J023237.90+723940.7.  The star is invisible in our J and H images, visible in our \Ks{} image, but is very bright in \mh{} image (leading to our designation for it as an MHO object).  MHO\,2953 and MHO\,2954 are faint, compact knots (Fig.\ \ref{FigMHO2953}).  MHO\,2959 is a short (5 arcsec long) filament with a knot on the southwestern end which lies 54 arcsec to the northeast of the position of IRAS\,F02277+7226.  Due to the numerous IRAS and WISE sources in the region, none of these four shocks can be positively associated with a source star in the region.

\subsubsection{The L1340\,C Mid-IR Cluster South:  MHO\,2955, MHO\,2956, MHO\,2958, MHO\,2963}
\label{MHO2955}

\begin{figure}[!htb]
\includegraphics[width=1.0\textwidth]{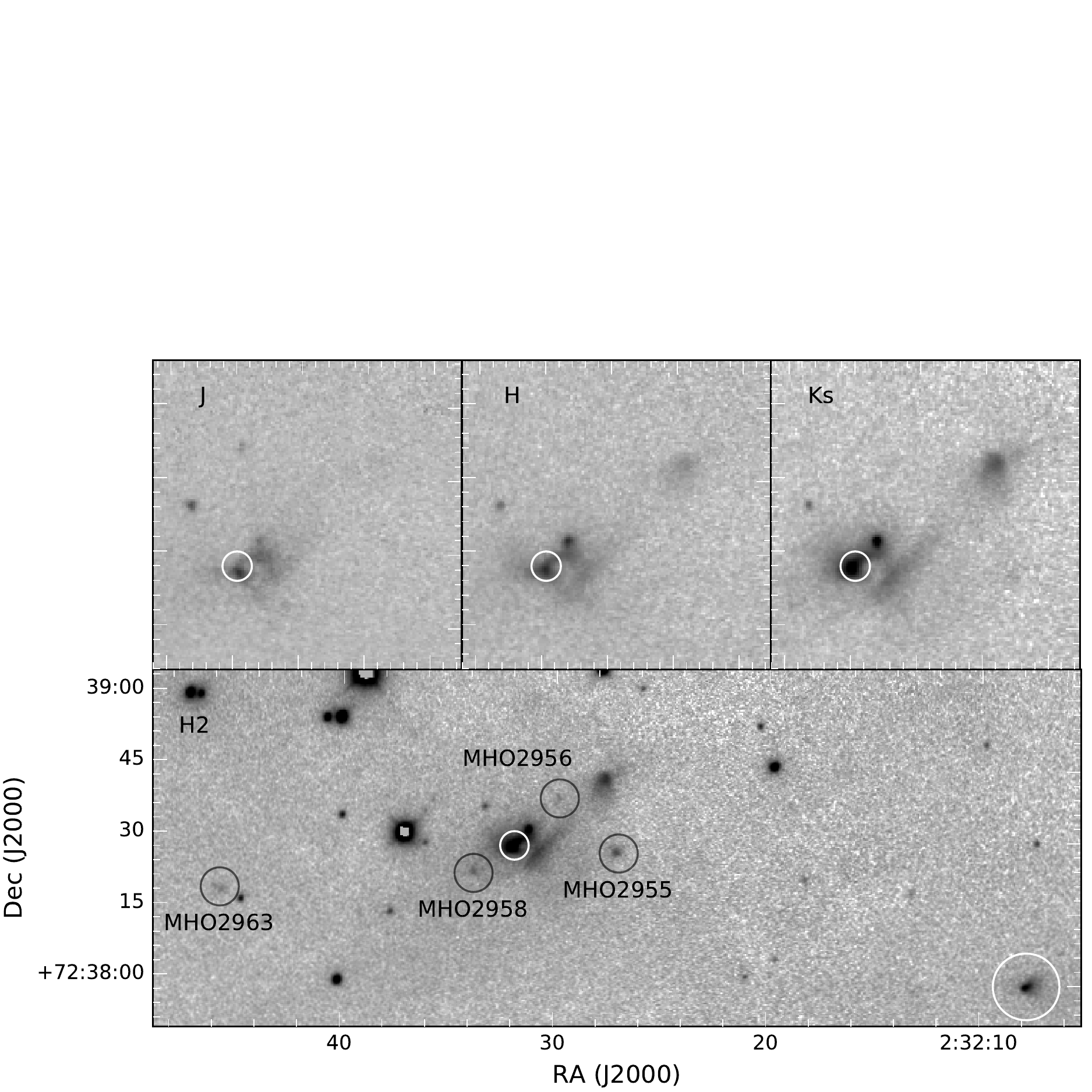}
\caption{An \mh{} image of the region discussed in \S\ref{MHO2955} containing MHO\,2955, MHO\,2956, MHO\,2958, \& MHO\,2963.  The upper panels show broadband images of the reflection nebulosity near the center of the \mh{} image.  The position of SSTSL2\,J023232.00+723827.5 is marked with a white circle.  The fan shaped reflection nebula discussed in the text is indicated by a large white circle in the lower right corner of the \mh{} image.}
\label{FigMHO2955}
\end{figure}

In the southern half of the cluster, roughly 45\arcsec{} south of IRAS\,F02279+7225 and IRAS\,02276+7225, we find a pair of stars separated by about 5 arcsec and surrounded by a reflection nebula (Fig.\ \ref{FigMHO2955}).  The southern star is coincident with the \cite{Kun2016IR} Class\,0/I source SSTSL2\,J023232.00+723827.5.  The reflection nebula is filamentary with one prominent filament offset from the stars to the southwest and oriented northwest/southeast.  It is suggestive of the brightened limb of a cavity oriented in that direction as the region between the filament and the stars is noticeably darker (especially in \Ks{}).  There is an additional bright knot of reflection nebulosity 19 arcseconds northwest of the northern star.  These stars are roughly coincident with the C3W NH$_3$ core of \cite{Kun2003}.

These stars are surrounded by three \mh{} knots (MHO\,2955, 2956, and 2958).  The brightest is MHO\,2955 which lies about 20 arcseconds west of the pair of stars.  MHO\,2956 is a faint, compact knot which lies about 9 arcseconds northwest of the northern star.  MHO\,2958 lies 10 arcseconds southeast of the southern star.  In addition, MHO\,2963 lies about 1\arcmin{} to the east.

Lastly, about 1.85 arcminutes to the west lies a fan shaped reflection nebula with the Class\,0/I source SSTSL2\,J023207.96+723759.3 at its apex.  Both the star and reflection nebula are visible in the \Ks{} and \mh{} images, but are very faint in J and H.  No \mh{} knots are detected in the vicinity of this nebula.

\subsubsection{The V1180\,Cas\,B Outflow:  MHO\,2964}
\label{MHO2964}

\begin{figure}[!htb]
\includegraphics[width=1.0\textwidth]{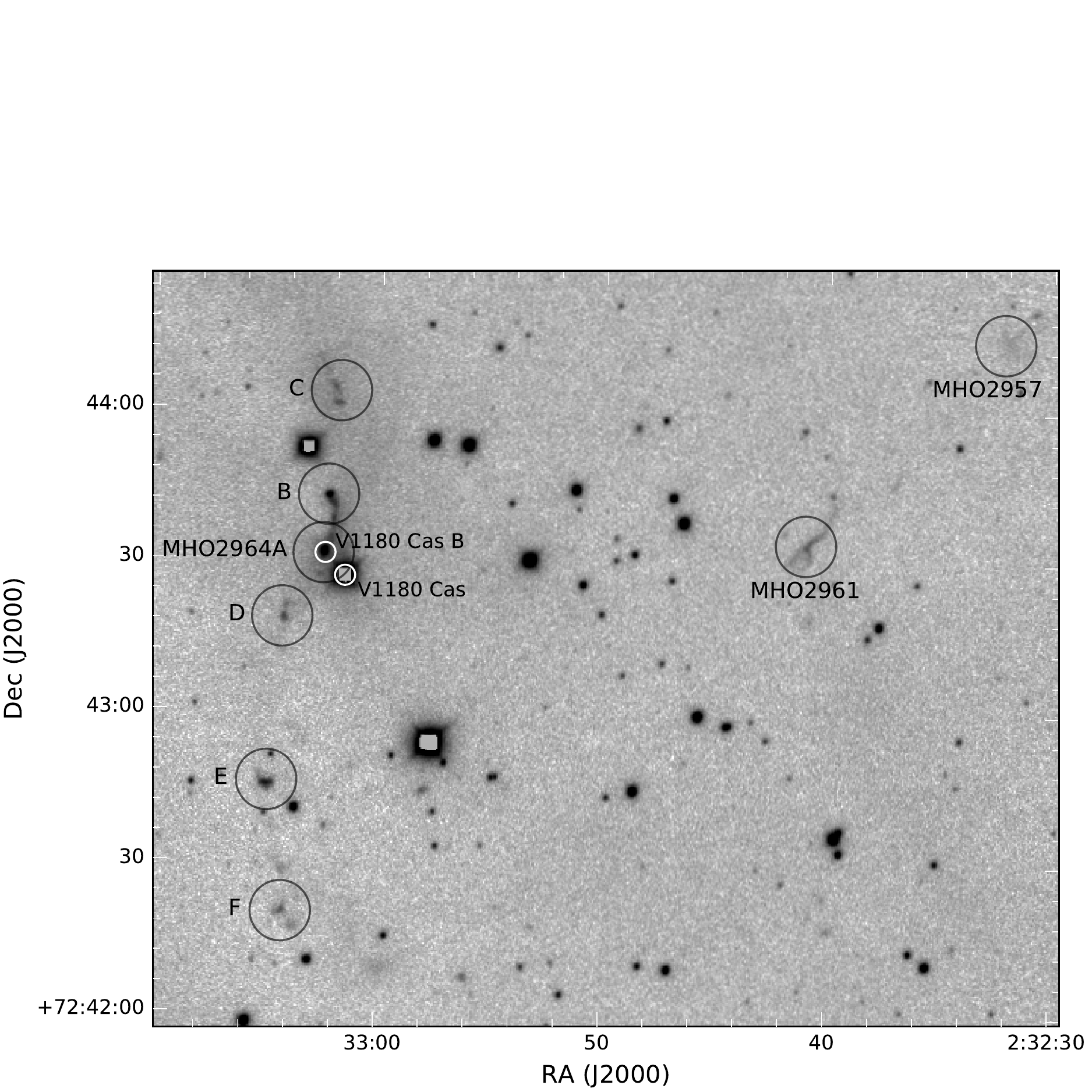}
\caption{An \mh{} image of the MHO\,2964 flow discussed in \S\ref{MHO2964}.  MHO\,2957 and MHO\,2961 are visible in the western half of the image.  V1180\,Cas and V1180\,Cas\,B are marked with white circles and labeled.}
\label{FigMHO2964}
\end{figure}

Roughly 5 arcminutes to the North of the group of IRAS sources lies a beautiful S-shaped outflow (MHO\,2964; Fig.\ \ref{FigMHO2964}).  The flow consists of 8 distinct knots stretching over 1.75 arcminutes along a north-south axis.

The flow axis passes near the position of two potential source stars, the brighter of which is the emission line star V1180\,Cas (emission line star 72 in Table 2 of \citealt{Kun2016Opt}).  V1180\,Cas coincides with SSTSL2\,J023301.52+724326.7, which \cite{Kun2016IR} identified as a Flat SED source.  \cite{Antoniucci2014} examined the region around these two stars and designated the fainter star V1180\,Cas\,B (which coincides with SSTSL2\,J023302.41+724331.2, a class\,0/I star catalogued by \citealt{Kun2016IR}).

Both \cite{Antoniucci2014} and \cite{Kun2016IR} detected MHO\,2964 (in \mh{} and Spitzer 4.5\,\micron{} bands respectively).  Our \mh{} images show that the MHO\,2964 chain of knots clearly passes through V1180\,Cas\,B (see Fig.\ \ref{FigMHO2964}), so we favor this star as the driving source as do \cite{Antoniucci2014} and \cite{Kun2016IR}.  The star is visible in our H and \Ks{} images, but is invisible in our J image.  

To the west of this source lie two additional \mh{} knots (MHO\,2961 and MHO\,2957; Fig.\ \ref{FigMHO2964}).  MHO\,2961 is a 14 arcsecond long filament oriented roughly northwest-southeast.  MHO\,2957 is a very faint, diffuse path of \mh{} emission.  Neither has a clear association with a source.

\subsubsection{MHO\,2962, MHO\,2985, MHO\,2966}
\label{MHO2962}

\begin{figure}[!htb]
\includegraphics[width=0.5\textwidth]{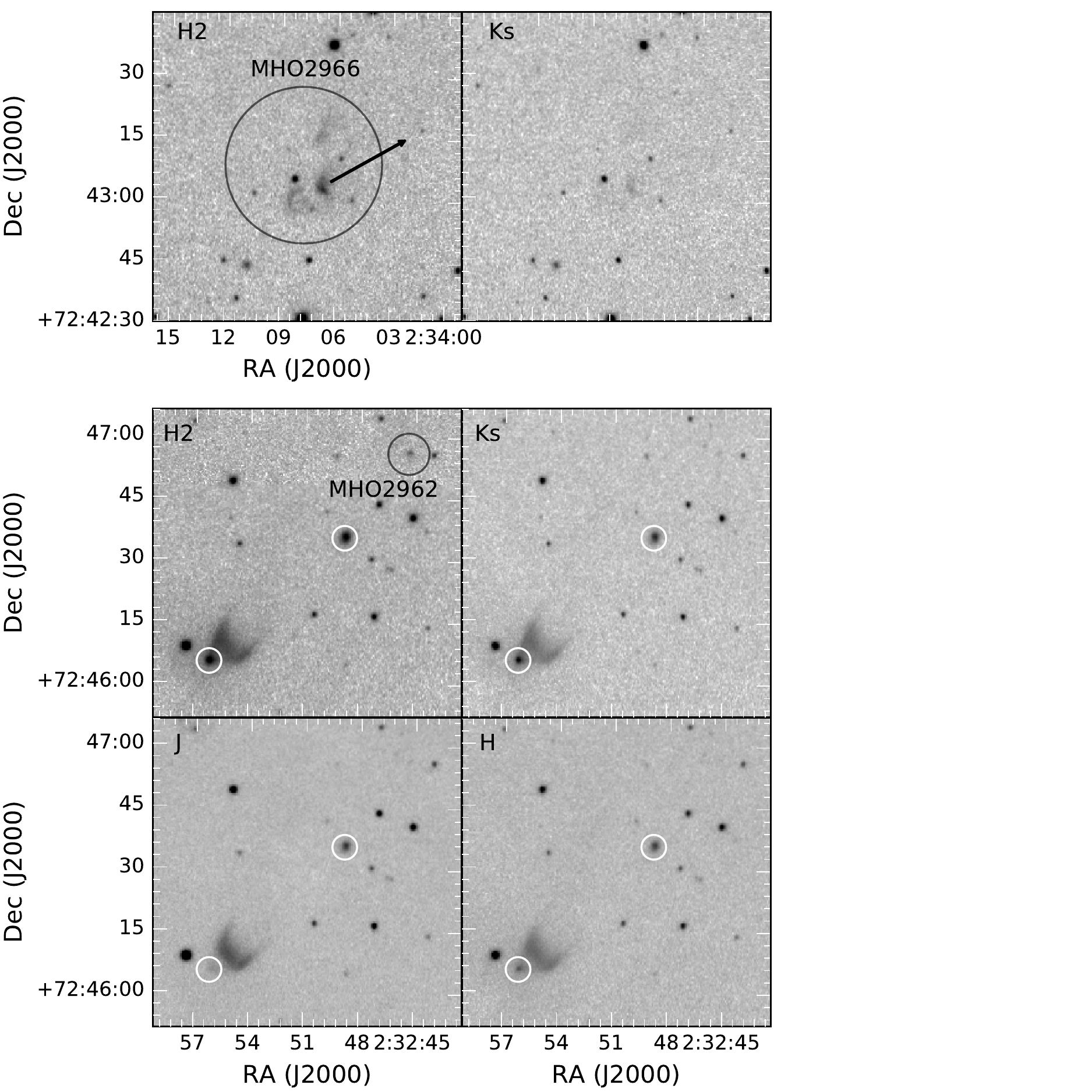}
\caption{Images of MHO\,2962 and MHO\,2966 discussed in \S\ref{MHO2962}.  The lower four panels show images of MHO\,2962 and its associated reflection nebula.  The positions of the two Class\,0/I sources discussed in the text are marked with white circles. The upper two panels show the MHO\,2966 shock which is likely a shock in the counterflow to MHO\,2962.  The arrow indicates the direction to the reflection nebula and to SSTSL2\,J023256.1+724605.3.}
\label{FigMHO2962}
\end{figure}

North of V1180 Cas lies a fan shaped reflection nebula (Fig.\ \ref{FigMHO2962}).  At the apex of the nebula is a star visible in our J, H, and \Ks{} images which corresponds to the Class\,0/I source SSTSL2\,J023256.14+724605.3 of \cite{Kun2016IR}.  To the northwest, 69 arcseconds away along the axis of the fan (PA $\sim$315 degrees) lies a compact \mh{} knot (MHO\,2962).  In addition, 6 arcminutes to the southeast, along a similar axis (PA $\sim$119 degrees), lies MHO\,2966 which is a cluster of 3 faint bow shocks facing back along the flow axis.

The star at the apex of the reflection nebula is one of the outbursting stars identified by \cite{Kun2014}.  They also identified two HH objects in narrowband \ha{} and \sii{}.  The SE object (using their nomenclature) is coincident with our MHO\,2962.  There is no conclusive counterpart in our \mh{} images to their NW object.

A second Class\,0/I source (STSL2\,J023248.83+724635.4) lies on a line drawn between the reflection nebula around SSTSL2\,J023256.1+724605.3 and MHO\,2962. The star appears slightly nebulous in our J, H, and \Ks{} images.

MHO\,2965 (Fig.\ \ref{FigL1340COverview}) is a \mh{} knot which lies 2.6 arcminutes North of the reflection nebula described in \S\ref{MHO2962} above.

\section{Discussion}\label{discussion}

\subsection{Comparison With Barnard\,1}\label{B1}

We have previously used a similar wide field near-IR survey of shocks to examine the Barnard\,1 region in Perseus \citep{Walawender2009}.  At roughly $\sim10^3$\,\Msun{}, the L1340 complex is roughly comparable in gas mass to Barnard\,1, but L1340 is divided up in to three sub-cores (discussed in \S\ref{L1340A}, \ref{L1340B}, \& \ref{L1340C} above), so it is morphologically different than the more cometary structure of Barnard\,1.  L1340 is also isolated unlike Barnard\,1 which is part of the larger Perseus Molecular Cloud complex which contains several other regions of intense star formation activity \citep{Walawender2005}.  Barnard\,1 also differs is that it is likely an example of triggered star formation.  \citet{Kirk2006} used the offset between submillimeter clumps and their parent extinction cores within Perseus to show evidence that star formation in Perseus (including that in Barnard\,1) has likely been triggered by soft UV radiation from a nearby B0.5 star, a hypothesis originally put forward by \citet{Walawender2004} based on the morphology of a single clump in the L1455 region of Perseus.

Outflow activity in L1340 appears to be somewhat more intense than in Barnard\,1, with roughly twice the number of shock complexes (24 in Barnard\,1 and 42 in L1340), but is less concentrated in that outflows in Barnard\,1 are dominated by those in a single central region with a few outliers while outflows in L1340 are distributed among the three cores.  

The very large, 3.7\,pc long outflow from SSTSL2\,J022907.88+724347.2 (discussed in \S\ref{MHO2925}) is more than twice as long as the most extensive flow found in Barnard\,1 and appears to be only one half of the flow as the Northeastern lobe was not detected in our images.  This gives it a dynamic age which is several times older than the oldest flow in Barnard\,1 under similar assumptions about outflow launch velocity.  This suggests that star formation in L1340 has been ongoing somewhat longer than in Barnard\,1, however the SSTSL2\,J022907.88+724347.2 outflow appears (in our J, H, and \Ks{} images and by comparison with the \cite{Kun2016Opt} extinction map) to propagate through a comparatively low extinction regime.  This could result in the flow not impacting as much ambient material, thus allowing it to travel at near-launch speeds over greater distances than flows which encounter surrounding material nearer their source protostar.

Despite the morphological differences (filamentary vs. clumpy, triggered vs. non-triggered), Barnard\,1 and L1340 appear to have similar populations of outflows with L1340's outflows likely being somewhat older.

\subsection{Properties of Outflow Driving Protostars}\label{properties}

In order to identify the candidate driving protostars for the outflows we described in \S\ref{Results}, we considered both the morphology of the individual shocks (for example, the bow shock morphology of MHO\,2925 discussed in \S\ref{MHO2925} was used to estimate the direction in which the source protostar must lie) and the positions of the individual shocks relative to a candidate source (for example, see the discussion of the source region for MHO\,2928 in \S\ref{MHO2928}).  The candidate sources we considered consisted of: \ha{} emission line stars identified in various previous works (primarily those in \citealt{Kun2016Opt}), IRAS sources identified as protostars in \cite{Kun1994}, the 45 Class\,0/I (8 of which they consider to be Class\,0 candidates) and 27 flat SED sources identified by \cite{Kun2016IR}.

Table \ref{TableFlows} contains a list of the 12 shock complexes which we feel clearly represent a single outflow structure and for which we have identified at least one strong candidate source protostar.  For only 7 of those 12 outflows can we unambiguously identify a single source protostar, for the other 5 outflows, we are limited to identifying 2-4 candidate source protostars.

For the subset of these candidate outflow sources which are also classified as Class\,0/I or Flat SED, we can compare the properties derived by \cite{Kun2016IR} of our candidate outflow source population with the properties of the Class\,0/I or Flat SED stars which are not outflow sources.  

\begin{figure}[!htb]
\includegraphics[width=1.0\textwidth]{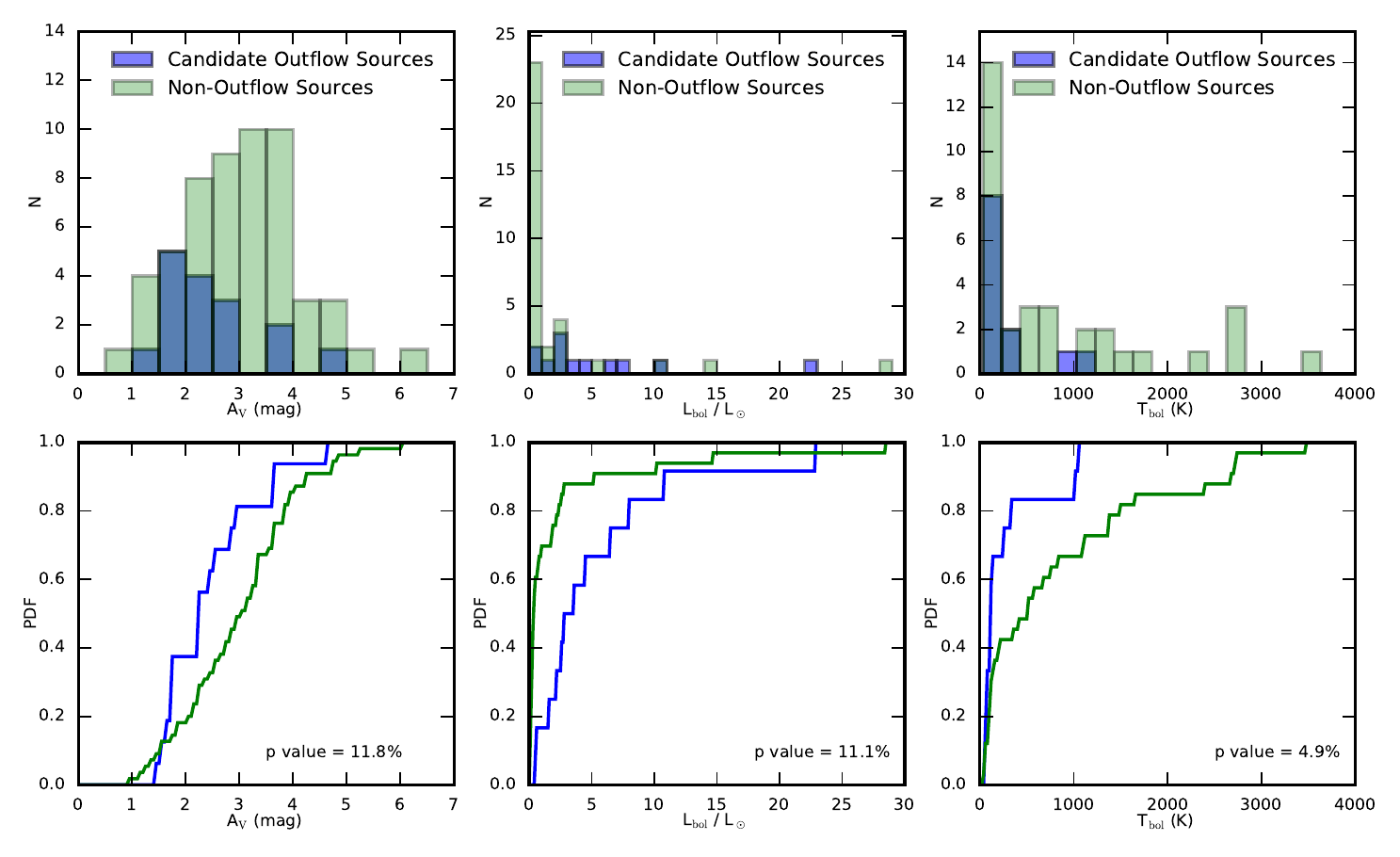}
\caption{A comparison of the A$_{V}$, T$_{bol}$, \& L$_{bol}$ values (from \cite{Kun2016IR} Tables 7 \& 8) for stars which are candidate outflow sources compared with the entire population of Class\,0/I and Flat SED protostars.  A histogram of the values for each population is shown in top row, while the cumulative probability distribution function for each property is shown in the bottom row.  The p-value quantifying the chances that these two populations are drawn from the same population is annotated in the lower right of each cumulative probability distribution function plot.}
\label{Properties}
\end{figure}

For each of the A$_{V}$, T$_{bol}$, and L$_{bol}$ properties, we compute a Kolmogorov-Smirnov statistic to determine whether the values for source candidates are drawn from the same distribution as the non-outflow source stars and determine a "p-value" which quantifies the chance that the source and non-source protostar properties are drawn from the same distribution.  Histograms for each property and the probability distribution function for obtaining each value are shown in Fig.\,\ref{Properties}.

While all three properties have p-values which suggest that the two populations differ, none of the p-values are small enough to confidently reject the hypothesis that they are drawn from the same population.

Candidate outflow sources appear (at the $\sim$88\% confidence level) to be drawn from a slightly lower A$_{V}$ population than are the non-outflow sources (see left column of Fig.\,\ref{Properties}).  This may be due, however, to selection effects.  For example, shocks may be obscured in the highest extinction regions, thus shocks near to the highest extinction protostars may not be detected by our survey and nearby shocks are the easiest to associate with a candidate protostar.

In the middle column of Fig.\,\ref{Properties} we see that the outflow source candidates lack the strong peak at low L$_{bol}$ which the non-source candidates do.  This is significant at a $\sim$89\% confidence level.  The candidate outflow sources are, on average, higher luminosity than the non-source candidates.

In the right hand column of Fig.\,\ref{Properties} we see that both distributions are clustered around small values of T$_{bol}$, however from the cumulative probability distribution function in the right hand panel, we see that proportionally, more outflow source candidates are clustered at small values than non-outflow candidates.  These two populations differ at the $\sim$95\% confidence level.  The bolometric temperature is a proxy for the evolutionary development of a protostar \citep{MyersLadd1993}.  One would expect outflow driving sources to be, on average, in the earlier stages of protostellar evolution and have lower bolometric temperatures, just as this result suggests.

We have also run the analysis above using only 7 of the 12 outflow driving protostars, selecting only those 7 for which there was an unambiguous identification of the source.  The statistical significance of this analysis with only 7 sources is limited, but the qualitative results are similar:  the p-value for A$_{V}$ increases to $\sim$33\% (making a selection effect less significant), while the p-value for T$_{bol}$ and L$_{bol}$ decrease to $\sim$1\% and $\sim$7\% respectively (nominally making the effect more significant).

\begin{figure}[!htb]
\includegraphics[width=0.5\textwidth]{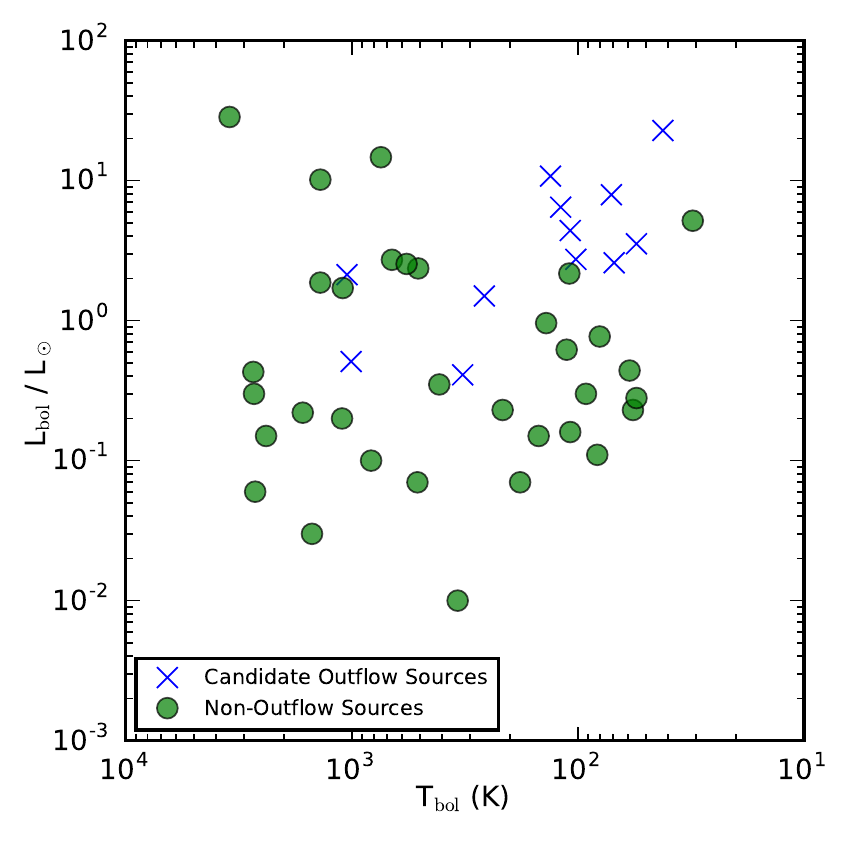}
\caption{An L$_{bol}$ vs. T$_{bol}$ plot for the candidate outflow sources (marked with blue x symbols) and the non-outflow sources (marked with filled green circles).}
\label{FigLbolTbol}
\end{figure}

We can also examine where outflow source candidates lie in a L$_{bol}$ vs. T$_{bol}$ plot in comparison to the non-outflow candidates.  This is analogous to \cite{Kun2016IR} Fig.\,12, but only including Class\,0/I and flat SED sources.  This is shown in Fig.\,\ref{FigLbolTbol} where the outflow source candidates can be seen to be biased toward the low T$_{bol}$ and high L$_{bol}$ side of the overall population of sources as would be expected based on the behaviors seen in Fig.\,\ref{Properties} above.

We also examined the source properties as a function of outflow length, but no correlation with any of A$_{V}$, T$_{bol}$, or L$_{bol}$ was apparent.

\section{Summary}\label{summary}

We have found 42 distinct shock complexes (MHO\,2925-2966) in the L1340 region.  Of those 42 shock complexes, we were able to link 17 of them in to 12 distinct outflows with candidate source stars (Table \ref{TableFlows}).  Of those 12 flows, 4 lie in L1340\,A, 5 in L1340\,B, and 3 in L1340\,C.  Six flows are longer than 1\,pc in length at the assumed distance to L1340 of 825\,pc.

The shocks which we were not able to link in to coherent flows tend to be concentrated in groups near the centers of the three regions.  In L1340\,A, 5 unassociated shocks lie within 1.7\arcmin{} of the RNO7 cluster (\S\ref{RNO7Region}).  In L1340\,B, 3 unassociated shocks lie within 1.5\arcmin{} of RNO8 (\S\ref{RNO8Region}).  Lastly, in L1340\,C, 8 unassociated shocks lie within a 1.5\arcmin{} radius of a group of IRAS sources (\S\ref{MHO2953} \& \S\ref{MHO2955}).  Thus 34 of our 42 shocks lie in identifiable outflows or in the more active (and thus confused) centers of the cloud.  Only 8 unassociated shocks lie in the outer areas of the cloud.

By combining our identification of candidate outflow producing protostars with the \cite{Kun2016IR} catalog of Class\,0/I and flat SED protostars in this region, we can begin to compare the properties of protostars which drive outflows with those protostars which do not appear to drive outflows.  With the work presented here, we have only a small sample which may suffer from selection effects, but we see some hints (at a level less than 2$\sigma$ confidence) that the properties (A$_{V}$, T$_{bol}$, \& L$_{bol}$) of outflow driving protostars differ from non-outflow driving protostars.  Outflow driving protostars appear to have slightly lower extinction (which could be a result of selection effects), have higher bolometric luminosity, and lower bolometric temperatures (which if we assume younger protostars are more likely to drive outflows is consistent with \citealt{MyersLadd1993} use of bolometric temperature as a proxy for a protostar's evolutionary state).

\acknowledgments

Based on observations obtained with WIRCam, a joint project of CFHT, Taiwan, Korea, Canada, France, at the Canada-France-Hawaii Telescope (CFHT) which is operated by the National Research Council (NRC) of Canada, the Institute National des Sciences de l'Univers of the Centre National de la Recherche Scientifique of France, and the University of Hawaii.

Based on observations obtained with the Apache Point Observatory 3.5-meter telescope, which is owned and operated by the Astrophysical Research Consortium.

This publication makes use of data products from the Wide-field Infrared Survey Explorer, which is a joint project of the University of California, Los Angeles, and the Jet Propulsion Laboratory/California Institute of Technology, funded by the National Aeronautics and Space Administration.

This research has made use of the VizieR catalogue access tool, CDS, Strasbourg, France.

The MHO catalogue is hosted by Liverpool John Moores University.

This research made use of Astropy, a community-developed core Python package for Astronomy \citep{astropy}.

G. W.-C. gratefully acknowledges support from the Brinson Foundation in aid of astrophysics research at the Adler Planetarium.

We would like to thank Adam Draginda, Rachael Zelman, and Mary Laychak (the CFHT queue observers) as well as Pierre Martin, Daniel Devost, and Todd Burdullis (the CFHT queue coordinators) who obtained our WIRCam data.

Finally, we would like to thank the University of Hawaii Time Allocation Committee for allocating the nights during which these observations were made. 

The authors also wish to recognize and acknowledge the very significant cultural role and reverence that the summit of Mauna Kea has always had within the indigenous Hawaiian community. We are fortunate to have the opportunity to conduct observations from this sacred mountain.

\begin{deluxetable}{lccl}
\tablecolumns{4}
\tabletypesize{\scriptsize}
\tablecaption{MHOs in L1340.}
\tablehead{ \colhead{MHO Designation} & \colhead{RA (J2000)} & \colhead{Dec (J2000)} & \colhead{Region} }
\startdata
MHO2925A &  2:26:21.0 & 72:34:37.1 &     L1340A \\
   2925B &  2:26:19.9 & 72:35:08.3 &     L1340A \\
   2925C &  2:26:27.8 & 72:34:57.3 &     L1340A \\
   2925D &  2:26:31.2 & 72:35:41.8 &     L1340A \\
   2925E &  2:26:36.1 & 72:35:12.7 &     L1340A \\
   2925F &  2:26:39.0 & 72:35:26.0 &     L1340A \\
   2925G &  2:26:39.5 & 72:36:18.2 &     L1340A \\
   2925H &  2:26:45.5 & 72:35:41.9 &     L1340A \\
MHO2926A &  2:27:33.5 & 72:34:08.4 &    L1340A \\
   2926B &  2:27:29.6 & 72:34:15.1 &    L1340A \\
 MHO2927 &   2:27:39.0 & 72:35:28.3 &    L1340A \\
MHO2928A &  2:27:56.4 & 72:35:58.7 &     L1340A \\
   2928B &  2:27:59.1 & 72:35:53.9 &     L1340A \\
   2928C &  2:27:59.8 & 72:35:57.9 &    L1340A \\
   2928D &  2:28:00.3 & 72:35:55.7 &    L1340A \\
   2928E &  2:28:06.2 & 72:35:29.2 &     L1340A \\
   2928F &  2:28:15.2 & 72:35:09.3 &     L1340A \\
   2928G &  2:28:22.6 & 72:34:55.6 &     L1340A \\
   2928H &  2:28:24.9 & 72:34:47.5 &     L1340A \\
   2928I &  2:28:28.5 & 72:34:56.0 &     L1340A \\
   2928J &  2:28:29.7 & 72:34:36.2 &     L1340A \\
   2928K &  2:28:35.6 & 72:34:33.6 &     L1340A \\
   2928L &  2:28:39.6 & 72:34:25.5 &     L1340A \\
   2928M &  2:28:42.0 & 72:34:17.1 &     L1340A \\
   2928N &  2:28:53.4 & 72:34:11.5 &     L1340A \\
   2928O &  2:28:56.9 & 72:34:04.0 &     L1340A \\
   2928P &  2:29:06.4 & 72:34:07.8 &     L1340A \\
 MHO2929 &   2:27:59.2 & 72:38:21.2 &     L1340A \\
 MHO2930 &   2:28:08.9 & 72:36:28.4 &     L1340A \\
 MHO2931 &   2:28:15.0 & 72:36:53.7 &     L1340A \\
MHO2932A &  2:28:15.6 & 72:37:45.3 &     L1340A \\
   2932B &  2:28:20.5 & 72:37:41.9 &     L1340A \\
 MHO2933 &   2:28:20.4 & 72:39:24.6 &     L1340A \\
 MHO2934 &   2:28:22.4 & 72:38:56.5 &    L1340A \\
 MHO2935 &   2:28:24.9 & 72:35:24.7 &    L1340A \\
 MHO2936 &   2:28:53.1 & 72:36:12.1 &    L1340A \\
MHO2937A &  2:29:07.2 & 72:43:42.3 &     L1340A \\
   2937B &  2:29:08.6 & 72:43:49.8 &     L1340A \\
MHO2938A &  2:29:29.8 & 72:37:34.6 &    L1340A \\
   2938B &  2:29:31.6 & 72:37:41.2 &    L1340A \\
MHO2939A &  2:29:43.1 & 72:43:53.9 &     L1340A \\
   2939B &  2:29:42.4 & 72:44:32.4 &     L1340A \\
   2939C &  2:29:37.8 & 72:44:52.8 &     L1340A \\
   2939D &  2:29:39.9 & 72:45:05.6 &     L1340A \\
   2939E &  2:29:40.4 & 72:45:13.1 &     L1340A \\
   2939F &  2:29:35.2 & 72:46:16.0 &     L1340A \\
   2939G &  2:29:44.3 & 72:43:45.8 &     L1340A \\
   2939H &  2:29:45.4 & 72:43:37.2 &     L1340A \\
   2939I &  2:29:44.9 & 72:42:15.6 &     L1340A \\
 MHO2940 &   2:30:06.8 & 72:39:54.1 &    L1340A \\
 MHO2941 &   2:26:42.0 & 72:54:24.4 &    L1340B \\
MHO2942A &  2:27:54.9 & 72:59:25.1 &    L1340B \\
   2942B &  2:27:49.0 & 72:59:30.9 &    L1340B \\
   2942C &  2:27:45.4 & 72:59:29.8 &    L1340B \\
   2942D &  2:27:26.4 & 72:59:27.4 &    L1340B \\
   2942E &  2:28:08.5 & 72:59:02.1 &    L1340B \\
   2942F &  2:28:17.1 & 72:58:45.3 &    L1340B \\
   2942G &  2:28:23.0 & 72:58:28.1 &    L1340B \\
   2942H &  2:28:23.8 & 72:58:33.8 &    L1340B \\
   2942I &  2:28:26.9 & 72:58:28.2 &    L1340B \\
   2942J &  2:29:12.9 & 72:57:56.0 &    L1340B \\
   2942K &  2:29:13.5 & 72:57:47.9 &    L1340B \\
   2942L &  2:29:18.1 & 72:57:44.1 &    L1340B \\
   2942M &  2:29:27.7 & 72:57:47.8 &    L1340B \\
MHO2943A &  2:27:55.7 & 73:03:53.2 &    L1340B \\
   2943B &  2:27:51.6 & 73:03:54.4 &    L1340B \\
   2943C &  2:28:00.1 & 73:03:54.6 &    L1340B \\
   2943D &  2:28:04.1 & 73:03:39.1 &    L1340B \\
 MHO2944 &   2:29:42.9 & 72:54:21.3 &    L1340B \\
 MHO2945 &   2:30:16.6 & 72:59:05.2 &    L1340B \\
MHO2946A &  2:30:02.8 & 73:02:51.9 &    L1340B \\
   2946B &  2:29:58.5 & 73:03:05.5 &    L1340B \\
   2946C &  2:29:30.2 & 73:03:32.2 &    L1340B \\
   2946D &  2:30:18.7 & 73:02:45.5 &    L1340B \\
   2946E &  2:30:31.2 & 73:02:29.4 &    L1340B \\
 MHO2947 &   2:30:44.9 & 73:02:51.0 &    L1340B \\
 MHO2948 &   2:31:13.3 & 73:02:37.1 &    L1340B \\
 MHO2949 &   2:30:47.6 & 72:59:30.2 &    L1340B \\
 MHO2950 &   2:30:53.7 & 72:59:07.8 &    L1340B \\
 MHO2951 &   2:32:14.3 & 72:56:10.2 &    L1340B \\
 MHO2952 &   2:31:30.4 & 72:39:42.5 &    L1340C \\
 MHO2953 &   2:32:21.7 & 72:40:03.1 &    L1340C \\
 MHO2954 &   2:32:22.1 & 72:39:49.0 &    L1340C \\
 MHO2955 &   2:32:27.0 & 72:38:26.7 &    L1340C \\
 MHO2956 &   2:32:29.8 & 72:38:38.1 &    L1340C \\
 MHO2957 &   2:32:32.2 & 72:44:14.1 &    L1340C \\
 MHO2958 &   2:32:33.8 & 72:38:22.2 &    L1340C \\
 MHO2959 &   2:32:35.1 & 72:40:32.1 &    L1340C \\
 MHO2960 &   2:32:37.9 & 72:39:41.0 &    L1340C \\
 MHO2961 &   2:32:41.0 & 72:43:33.7 &    L1340C \\
 MHO2962 &   2:32:45.4 & 72:46:56.1 &    L1340C \\
 MHO2963 &   2:32:45.7 & 72:38:18.6 &    L1340C \\
MHO2964A &  2:33:02.5 & 72:43:31.2 &    L1340C \\
   2964B &  2:33:02.3 & 72:43:42.8 &    L1340C \\
   2964C &  2:33:01.8 & 72:44:03.4 &    L1340C \\
   2964D &  2:33:04.3 & 72:43:18.5 &    L1340C \\
   2964E &  2:33:04.9 & 72:42:46.0 &    L1340C \\
   2964F &  2:33:04.2 & 72:42:20.0 &    L1340C \\
 MHO2965 &   2:33:09.6 & 72:48:33.9 &    L1340C \\
MHO2966A &  2:34:06.7 & 72:43:03.5 &    L1340C \\
   2966B &  2:34:06.8 & 72:43:15.5 &    L1340C \\
   2966C &  2:34:08.2 & 72:42:59.7 &    L1340C \\
\enddata
\label{TableShocks}
\end{deluxetable}

\begin{deluxetable}{llllll}
\tablecolumns{6}
\tabletypesize{\scriptsize}
\rotate
\tablecaption{Outflows in L1340.}
\tablehead{ \colhead{Designation} & \colhead{Length (\arcmin{}, pc)} & \colhead{NH$_3$ Core} & \colhead{Source Candidates} & \colhead{Class} & \colhead{Shock Components} }
\startdata
L1340A Flow 1 & 15.5\arcmin{}, 3.7\,pc  & A3      & SSTSL2\,J022907.88+724347.2 & Flat SED   & MHO\,2925, 2937, HH\,487 \\[5mm]
L1340A Flow 2 &  5.7\arcmin{}, 1.4\,pc  & A1      & SSTSL2\,J022818.51+723506.2 & Class\,0/I & MHO\,2928 \\
 \nodata      & \nodata                 & \nodata & SSTSL2\,J022820.81+723500.5 & Class\,0   & \nodata \\[5mm]
L1340A Flow 3 &  9.9\arcmin{}, 2.4\,pc  & A1      & SSTSL2\,J022844.40+723533.5 & Class\,0/I & MHO\,2926, 2935, 2936, 2938, 2940, HH\,488\,D, HH\,672 \\
\nodata       & \nodata                 & \nodata & SSTSL2\,J022842.57+723544.3 & Class\,0/I & \nodata \\
\nodata       & \nodata                 & \nodata & WISE\,J022817.97+723517.5   & \nodata    & \nodata \\[5mm]
L1340A Flow 4 &  4.0\arcmin{}, 0.7\,pc  & A4      & SSTSL2\,J022943.01+724359.6 & Class\,0/I & MHO\,2939 \\
\nodata       & \nodata                 & \nodata & SSTSL2\,J022943.64+724358.6 & Class\,0/I & \nodata \\[5mm]
L1340B Flow 1 &  9.1\arcmin{}, 2.2\,pc  & \nodata & SSTSL2\,J022808.60+725904.5 & Class\,0   & MHO\,2942 \\[5mm]
L1340B Flow 2 & 13.3\arcmin{}, 3.2\,pc  & \nodata & SSTSL2\,J022931.98+725912.4 & Class\,0   & MHO\,2941 \\[5mm]
L1340B Flow 3 &  1.0\arcmin{}, 0.24\,pc & \nodata & SSTSL2\,J022756.91+730354.4 & Class\,0/I & MHO\,2943 \\[5mm]
L1340B Flow 4 &  4.5\arcmin{}, 1.08\,pc & B1      & SSTSL2\,J022955.10+730309.1 & Class\,0/I & MHO\,2946 \\
\nodata       & \nodata                 & \nodata & SSTSL2\,J023020.61+730233.7 & Flat SED   & \nodata \\[5mm]
L1340B Flow 5 &  0.3\arcmin{}, 0.07\,pc & B2      & SSTSL2\,J023042.36+730305.1 & Class\,0/I & MHO\,2947 \\[5mm]
L1340C Flow 1 & 0.65\arcmin{}, 0.16\,pc & \nodata & WISE J023127.19+724015.9    & \nodata    & MHO\,2952 \\
\nodata       & \nodata                 & \nodata & SSTSL2\,J023127.34+724012.9 & Class\,0/I & \nodata \\[5mm]
L1340C Flow 2 & 1.75\arcmin{}, 0.30\,pc & \nodata & SSTSL2\,J023302.41+724331.2 & Class\,0/I & MHO\,2964 \\[5mm]
L1340C Flow 3 & 1.15\arcmin{}, 0.20\,pc & \nodata & SSTSL2\,J023256.14+724605.3 & Class\,0   & MHO\,2962 \\
\enddata
\label{TableFlows}
\end{deluxetable}

{\it Facilities:}  \facility{CFHT (WIRCAM)}, \facility{ARC (NICFPS)}, \facility{WISE}

\bibliography{L1340_refs}

\end{document}